\documentclass[11pt]{article}
\usepackage[a4paper,margin=1in]{geometry}
\usepackage{amsmath,amssymb,amsthm,mathtools}
\usepackage{microtype,booktabs,xcolor}
\usepackage{cite}
\usepackage{tabularx}
\usepackage{algorithm}
\usepackage[noend]{algpseudocode}
\usepackage{tikz}
\usetikzlibrary{arrows.meta,positioning,calc,decorations.pathreplacing}
\usepackage{authblk}
\usepackage[colorlinks=true,linkcolor=blue!55!black,citecolor=blue!55!black,urlcolor=blue!55!black]{hyperref}

\newtheorem{theorem}{Theorem}[section]
\newtheorem{lemma}[theorem]{Lemma}

\newtheorem{corollary}[theorem]{Corollary}

\theoremstyle{definition}

\newcommand{\E}{\mathbb E}
\newcommand{\OPT}{\operatorname{OPT}}
\newcommand{\ALG}{\operatorname{ALG}}
\newcommand{\RoE}{\operatorname{RoE}}

\newcommand{\cost}{\operatorname{cost}}

\usepackage{tcolorbox}
\tcbuselibrary{skins,breakable}
\usepackage{caption}
\definecolor{PaperBlue}{HTML}{315B88}
\definecolor{PaperTeal}{HTML}{21776F}
\definecolor{PaperAmber}{HTML}{A46422}
\definecolor{PaperPurple}{HTML}{705A87}
\definecolor{PaperInk}{HTML}{243449}
\hypersetup{linkcolor=PaperBlue,citecolor=PaperTeal,urlcolor=PaperBlue}
\tcbset{paperstatement/.style={enhanced,sharp corners,
  boxrule=.45pt,left=5pt,right=5pt,top=3pt,bottom=3pt,
  before skip=8pt,after skip=8pt}}
\tcolorboxenvironment{theorem}{paperstatement,
  colback=PaperTeal!5,colframe=PaperTeal!75!black}
\tcolorboxenvironment{lemma}{paperstatement,frame hidden,
  borderline west={1.5pt}{0pt}{PaperBlue},colback=PaperBlue!3}
\tcolorboxenvironment{proposition}{paperstatement,frame hidden,
  borderline west={1.5pt}{0pt}{PaperPurple},colback=PaperPurple!4}
\tcolorboxenvironment{corollary}{paperstatement,
  colback=PaperTeal!3,colframe=PaperTeal!50}
\tcolorboxenvironment{definition}{paperstatement,frame hidden,
  borderline west={1.5pt}{0pt}{PaperAmber},colback=PaperAmber!4}
\title{Stochastic Aggregation with Prescribed Resources\\and Additive Delay}
\author{Tianhang Lu\thanks{The first two authors contribute equally.}, Runtian Ren, Shengcai Liu}
\affil{Guangdong Provincial Key Laboratory of Brain-Inspired Intelligent Computation,\\
Department of Computer Science and Engineering,\\
Southern University of Science and Technology, Shenzhen 518055, China\\
liusc3@sustech.edu.cn}
\date{}

\begin{document}
\maketitle
\begin{abstract}
We study online aggregation with prescribed resources and additive delay.
In this problem, we are given a finite set of resources with purchase
costs and a collection of request types, each specifying the resources
it requires. Requests arrive over time, and each request can be served
immediately or postponed to a later time. A service may serve several
pending requests together by purchasing every resource required by at
least one of them, paying for each purchased resource only once.
These purchases serve only the current batch; resources needed in a
later service must be purchased again. A request served $t$ units of
time after its arrival incurs delay cost $t$. The objective is to serve
all requests while minimizing the sum of resource purchase costs and
individual delay costs.

We consider a stochastic version of this problem in which requests of
each type follow an independent Poisson process of known rate.
An online algorithm knows these rates but observes requests only upon
arrival. We compare its expected cost with that of an offline optimum
that knows the entire realized input and may choose every batch freely.
A resource--scale linear program supplies both an offline lower
certificate and synchronized service periods. This yields a deterministic
online algorithm with polynomial-time preprocessing on rational input
and a long-run ratio of expectations below $2.113$. The guarantee applies
in particular to rooted multi-level aggregation and prescribed-pair
aggregation on trees. Besides, for Poisson TCP acknowledgment, we further characterize the optimal long-run ratio of expectations at every load, showing that its worst-case value is approximately 1.409 and is attained by a deterministic online policy.
\end{abstract}

\section{Introduction}\label{sec:intro}
Consider a service system in which each order requires several shared
operations. An order may need transmission along a network path, access
to a group of machines, or the replenishment of several items. Handling
orders immediately repeats these purchases. Waiting allows orders with
common requirements to share their cost, but every waiting customer
incurs a delay penalty. Even when the required resources are known in
advance, deciding which orders to combine remains an online problem.

Rooted multi-level aggregation is a familiar example: a request at a
vertex needs every edge on its path to the root. Pair demands on a tree
instead need the path between two specified endpoints. More general
requirements need not come from paths or have a hierarchy. Three types
may require $\{e_1,e_2\}$, $\{e_2,e_3\}$, and $\{e_1,e_3\}$, respectively.
Each pair can share one purchase, but no single tree of nested
requirements explains all three possibilities. We call the general
model \emph{aggregation with prescribed resources}: a type specifies a
fixed resource set, and a batch pays the weight of the union of the
sets required by its requests.

Aggregation, replenishment, and related network design problems with
delay have received sustained attention in the adversarial model
\cite{bienkowski-et-al-2016,buchbinder-2017,carrasco-2018,
azar-touitou-2019,azar-touitou-2020,azar-setcover-2020,
touitou-lower-2021,mcm-2021,le-2023,azar-iluz-2026,ahmadian-2026}.
Their guarantees depend on the service model and on the information
available about delay. For example, the recent $2D$ bound for general
delay in rooted aggregation depends on tree depth $D$
\cite{ahmadian-2026}. Broader service systems also have superconstant
lower bounds \cite{touitou-lower-2021}. Such lower bounds must not be
transferred automatically to every tree model or every stochastic
arrival assumption.

A provider may know more than the worst-case model allows. Past demand
can indicate how often a resource is needed, even though it cannot
predict the next order. We isolate this information by assigning every
request type an independent Poisson process of known rate. The rates
are available before execution; the actual requests are revealed only
when they arrive. The comparison remains demanding: the offline optimum
knows the entire realized sequence and may choose every batch freely.
The question is therefore the following.
\begin{quote}\itshape
Can arrival rates coordinate arbitrary overlapping requirements well
enough to obtain a constant guarantee against the clairvoyant optimum?
\end{quote}

Mari, Paw\l{}owski, Ren, and Sankowski studied ratios of expectations
for stochastic matching with delays and rooted multi-level aggregation
\cite{mari-et-al-2023,mari-et-al-2024}. Their aggregation algorithm
combines immediate service of light demand with periodic service of
heavy demand, and their paper reports a general bound of $210$.
These works motivate the use of statistical information, but do not
coordinate arbitrary resource overlap. Our resource--scale LP handles
this overlap and gives a substantially smaller bound for the same
uncapacitated rooted problem. The algorithm and proof are independent
of the earlier GEN analysis; the present result does not validate or
repair that analysis.

\subsection{Our contributions}
We study independent Poisson arrivals of known rates, with each request
paying its waiting time. Our guarantees compare the expected cost of
an online algorithm with the expected cost of an offline optimum that
knows all future arrivals. We consider this ratio over increasingly
long horizons, including the full cost of completing every request.
Our results address two questions: how well statistical information
can coordinate many overlapping resource requirements, and how much
it can help even in the simplest one-resource problem.

\begin{itemize}
\item \textbf{A constant guarantee beyond tree aggregation.}
We give a deterministic online algorithm with a ratio below $2.111$
for arbitrary prescribed resource sets. Its guarantee is independent
of the number of resources, the number of request types, and how their
requirements overlap. With a small loss in the constant, we obtain a
ratio below $2.113$ with polynomial-time preprocessing for rational
costs and arrival rates.

This general model includes rooted multi-level aggregation: the
resources required by a request are the edges on its path to the root.
It also includes aggregation for prescribed pairs of vertices on a
tree. Both problems therefore inherit our guarantees. In particular,
we improve on the $210$ bound reported for stochastic rooted MLA to below
$2.111$, while extending the result to requirements with no tree
structure.

\item \textbf{The optimal guarantee for Poisson TCP acknowledgment.}
When all requests need the same resource, the problem becomes TCP
acknowledgment: one acknowledgment can serve several pending requests,
but each request pays for its own delay. For arbitrary arrival
sequences, the optimal randomized competitive ratio is
$e/(e-1)\approx1.582$ \cite{karlin-kenyon-randall-2001}.
We determine the best possible long-run ratio of expectations under
Poisson arrivals, obtaining a worst-case value of about $1.409$.
A deterministic online policy attains this value, and randomization
cannot improve it.

Our characterization is more detailed than a single worst-case bound.
For every positive arrival rate and acknowledgment cost, we determine
the optimal online cost rate and characterize the offline cost rate.
Together they give the optimal ratio for that particular load, which
can be substantially smaller than the worst-case value. We also
identify the hardest load. Thus the result quantifies exactly the
remaining advantage of knowing the future after the arrival rate is
already known. Since TCP acknowledgment is a special case of our
general model, it also supplies a lower bound for prescribed-resource
and tree aggregation.
\end{itemize}

Table~\ref{tab:results} summarizes these two main contributions and
their consequences for trees.

\begin{table}[t]
\centering\small
\renewcommand{\arraystretch}{1.2}
\begin{tabularx}{\textwidth}{@{}Xccc@{}}
\toprule
\textbf{Problem}&\multicolumn{2}{c}{\textbf{Upper bound}}&\textbf{Lower bound}\\
\cmidrule(lr){2-3}
&Finite prep.&Polynomial prep.&\\
\midrule
Arbitrary prescribed resources & $<2.111$ & $<2.113$ & $>1.408$\\
Rooted and pair aggregation on trees & $<2.111$ & $<2.113$ & $>1.408$\\
Poisson TCP acknowledgment & \multicolumn{3}{c}{Optimal ratio: about $1.409$}\\
\bottomrule
\end{tabularx}
\caption{Worst-case long-run ratios of expectations under known-rate
Poisson arrivals. All upper bounds are attained by deterministic online
policies. In the first two rows, the polynomial-time preprocessing
guarantee assumes rational costs and rates. The TCP result supplies the lower bounds in the first two
rows and also characterizes the optimal ratio at every load.}
\label{tab:results}
\end{table}

\subsection{Our techniques}
\paragraph{Coordinating purchases through arrival rates.}
The basic tradeoff is familiar. Serving requests immediately avoids
waiting but repeats purchases; collecting requests into batches saves
purchase costs but makes them wait. Arrival rates tell us how much
sharing we can expect from waiting. With several resources, however,
these decisions interact. Two types may benefit from sharing a costly
resource even though their other requirements are different. Choosing
when to serve each type in isolation can lose this saving, while
forcing every type to follow the same schedule can create excessive
delay.

Our approach chooses service frequencies jointly and coordinates the
resulting service times. Types can be served at different frequencies
while still meeting regularly to share purchases. The algorithm buys
only the resources needed by requests actually being served. The main
challenge is to show that these simple scheduling choices compete
with an offline optimum that is free to arrange its batches after
seeing the entire input.

\paragraph{Comparing both schedules through the same lower bound.}
We use a linear program that accounts for resource costs at different
time scales. Its lower-bound interpretation splits the original costs
among simpler one-resource problems, without counting any purchase or
any request's waiting cost more than once. Every offline schedule must
pay at least the sum of these simpler optima. Their Poisson analysis
therefore gives a lower bound even for an optimum that knows the future.

The same LP guides the online scheduling choices. This is the essential
connection: the lower bound and the algorithm account for the same
opportunities to share costs. We can consequently bound both purchase
and waiting costs without a loss depending on the size or structure of
the resource system. Keeping track of the occasions when there is
nothing to serve, and carefully choosing the parameters, gives the
stated constant. An efficient preprocessing procedure preserves the
guarantee with an arbitrarily small multiplicative loss.

\paragraph{Understanding the advantage of knowing the future.}
The one-resource problem supports the general analysis and is also
interesting in its own right. An online policy knows how frequently
requests arrive, but it cannot tell whether the next request is about
to arrive or is still far away. The offline optimum can use precisely
this information when deciding which requests to combine.

We determine the best cost achievable without this future information
and separately characterize the cost achievable with it. Comparing
the two gives the optimal ratio at each load and identifies the worst
load. This comparison accounts for all causal policies, including
randomized ones, and all offline batching choices. It thus establishes
the limit of what arrival-rate information can achieve in this basic
problem, in addition to supplying the one-resource estimates used in
the general algorithm.

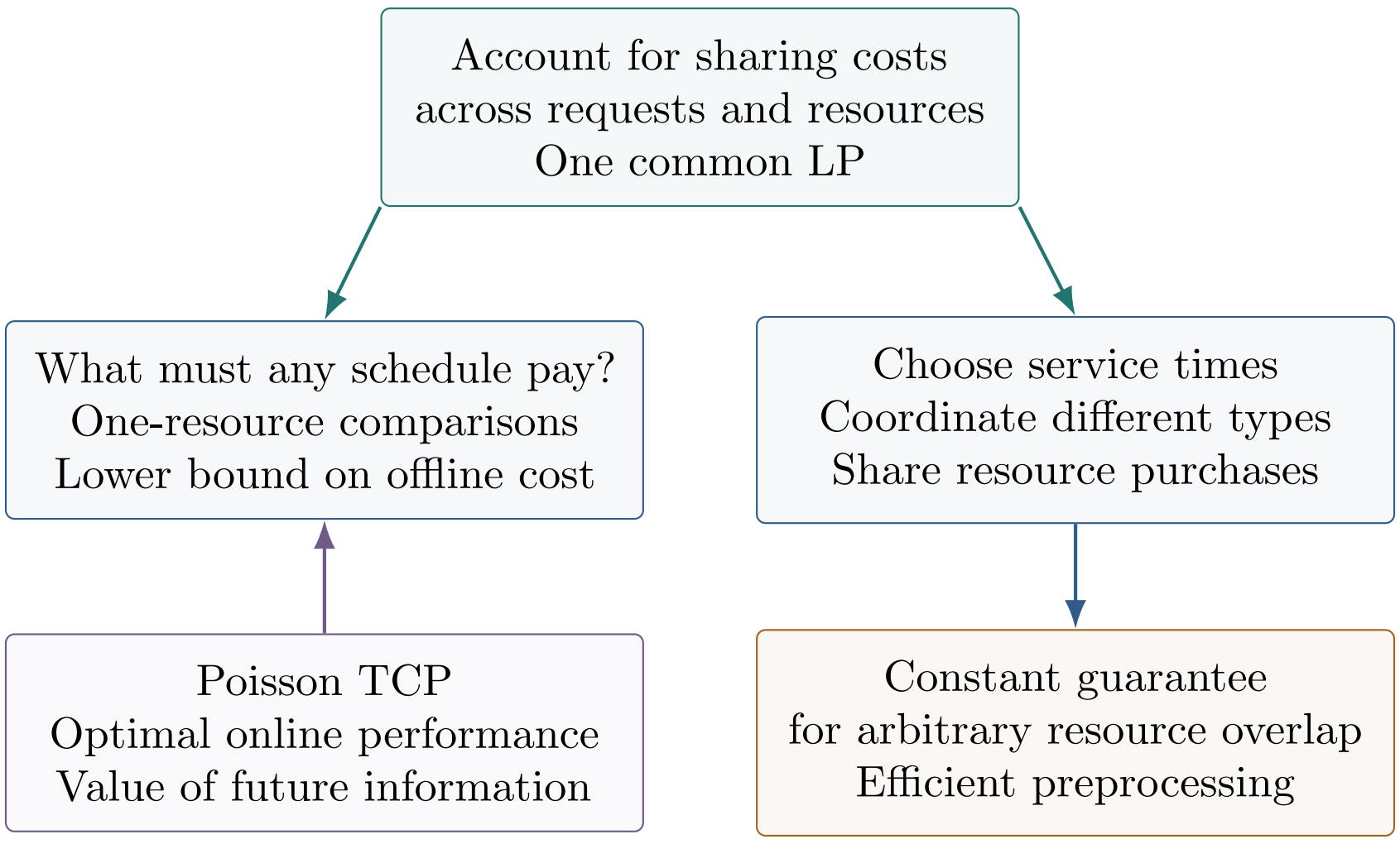
\begin{figure}[t]
\centering
\begin{tikzpicture}[font=\small,>=Latex,
 box/.style={draw,rounded corners=2pt,text width=4.6cm,align=center,
 inner sep=7pt,minimum height=1.4cm}]
\node[box,draw=PaperTeal,fill=PaperTeal!5] (budget) at (0,0)
 {Account for sharing costs\\across requests and resources\\One common LP};
\node[box,draw=PaperBlue,fill=PaperBlue!4] (lower) at (-3,-2.5)
 {What must any schedule pay?\\One-resource comparisons\\Lower bound on offline cost};
\node[box,draw=PaperBlue,fill=PaperBlue!4] (upper) at (3,-2.5)
 {Choose service times\\Coordinate different types\\Share resource purchases};
\node[box,draw=PaperAmber,fill=PaperAmber!5] (finite) at (3,-5)
 {Constant guarantee\\for arbitrary resource overlap\\Efficient preprocessing};
\node[box,draw=PaperPurple,fill=PaperPurple!4] (exact) at (-3,-5)
 {Poisson TCP\\Optimal online performance\\Value of future information};
\draw[->,thick,PaperTeal] (budget.south west)--(lower.north);
\draw[->,thick,PaperTeal] (budget.south east)--(upper.north);
\draw[->,thick,PaperBlue] (upper)--(finite);
\draw[->,thick,PaperPurple] (exact)--(lower);
\end{tikzpicture}
\caption{The LP gives both a lower bound on what any offline schedule
must pay and coordinated service periods for the online algorithm.
The Poisson TCP analysis supplies the one-resource comparisons and
an unavoidable lower bound for all causal policies.}
\label{fig:framework}
\end{figure}

\subsection{Related work}
\paragraph{Aggregation and replenishment.}
Rooted aggregation includes acknowledgement and joint replenishment as
shallow cases. Bienkowski and coauthors obtained an $O(D^4 2^D)$
guarantee for general delay on depth-$D$ trees
\cite{bienkowski-et-al-2016}. For deadlines, Buchbinder, Feldman, Naor,
and Talmon obtained $O(D)$ \cite{buchbinder-2017}, and McMahan obtained
$D$ \cite{mcm-2021}. Azar and Touitou improved the general-delay bound
to $O(D^2)$ \cite{azar-touitou-2019}; Ahmadian and coauthors' recent
preprint gives $2D$ by dual fitting \cite{ahmadian-2026}.
Azar and Iluz study nonmonotone delay penalties \cite{azar-iluz-2026},
whereas our penalties are known linear functions.

\paragraph{Set-based service and general frameworks.}
Carrasco, Pruhs, Stein, and Verschae study online set aggregation and
give an $O(\log m)$ bound in the number of requests, with a matching
randomized lower bound in their model \cite{carrasco-2018}.
Azar and Touitou's frameworks handle several network design problems
with deadlines or delay \cite{azar-touitou-2019,azar-touitou-2020}.
The latter uses offline optimization primitives, including
prize-collecting versions for delay. Set cover with delay and
nonclairvoyant frameworks address different service choices or
information assumptions \cite{azar-setcover-2020,touitou-nonclair-2023}.
Touitou proves lower bounds for several broader network models
\cite{touitou-lower-2021}. Our union cost is a specified coverage
function, and independent known-rate arrivals replace adversarial input.

\paragraph{Statistical information.}
The stochastic matching and rooted aggregation papers
\cite{mari-et-al-2023,mari-et-al-2024} are our closest predecessors in
arrival model and performance criterion. We preserve their distinction
between a ratio of expectations and an expected samplewise ratio.
Information about rates is also distinct from knowing a request's future
delay function, as studied by Le, Umboh, and Xie \cite{le-2023}.
Every delay function here is known; its future request arrivals are not.

\subsection{Organization}
Section~\ref{sec:model} defines the model and recalls the basic Poisson
facts. Section~\ref{sec:scalar-story} explains the optimal TCP guarantee,
and Section~\ref{sec:resource-story} develops the general-resource
algorithm and the comparisons behind its ratio. Together these
sections give the main results and their proof ideas.
Sections~\ref{sec:technical-tools}--\ref{ag:sec:pair-additive} provide the
technical tools and complete proofs. Appendix~\ref{app:scalar-certificates}
gives the arithmetic certificates supporting the numerical constants.

\section{Model and basic Poisson facts}\label{sec:model}
\subsection{Requests, services, and costs}
Let $E$ be a finite set of resources. Resource $e$ has purchase cost
$w_e>0$, and $w(S)=\sum_{e\in S}w_e$ for $S\subseteq E$.
There is a finite set $I$ of request types. Each type $i$ has a
prescribed nonempty resource set $R_i\subseteq E$ and arrival rate
$\lambda_i>0$. Zero-rate types and unused resources may be discarded;
requests with no positive-cost requirements can be served for free.
We consider nontrivial instances below.

A request is one occurrence of a type. Write $x(r)$ for the type of
request $r$, $a(r)$ for its arrival time, and $t(r)\ge a(r)$ for its
service time. A service chooses a batch $B$ of pending requests,
purchases the union of their required resources, and serves them
instantaneously. Its purchase cost is
\[
 w\left(\bigcup_{r\in B}R_{x(r)}\right).
\]
Each resource is paid for once in this batch, regardless of how many
requests need it. A later batch must purchase its resources again.
Every request receives all its required resources in one service and
incurs waiting cost $t(r)-a(r)$. The objective is the total purchase
cost plus the sum of these individual waiting costs.

Services have no duration or capacity restriction, and there is no
additional fee per service. An online algorithm may use the known
instance and requests observed so far, but not future arrivals.
Arrivals at a service time are observed first; simultaneous services
are processed in execution order. An offline optimum knows the entire
realized input and may choose any feasible batches and service times.

\subsection{Poisson arrivals and the performance criterion}
Each type follows an independent homogeneous Poisson process of its
known rate. Equivalently, successive interarrival times for type $i$
are independent exponential variables of mean $1/\lambda_i$.
The rates are known before execution; the realized arrival times are
revealed only as requests arrive.

For a horizon $T>0$, let $I_T$ be the arrivals in $[0,T]$. The algorithm
receives no signal that the input has ended. It must eventually serve
all requests, and $\ALG_T$ includes the entire cost of doing so, even
when some services occur after $T$. Let $\OPT_T$ be the minimum cost
of completing the same input. For a fixed instance we measure
\begin{equation}\label{eq:roe}
 \RoE(\ALG)=\limsup_{T\to\infty}
 \frac{\E[\ALG_T]}{\E[\OPT_T]}.
\end{equation}
Thus we compare expected total costs over increasingly long horizons.
This is a ratio of expectations, rather than the expectation of a
samplewise ratio. Expectations include independent algorithmic
randomness when relevant. We write $O_I(1)$ for a finite constant
that may depend on the fixed instance but not on $T$.

\subsection{The basic facts we will use}
For a rate-$\lambda$ process, an interval of fixed length $p$ contains
$\lambda p$ requests in expectation and is empty with probability
$e^{-\lambda p}$. Counts in disjoint intervals are independent.
Combining independent Poisson streams adds their rates; independently
retaining each arrival with probability $u$ gives a Poisson stream
of rate $u\lambda$. These facts let us group request types or extract
simpler comparison instances without losing the arrival model.

Conditional on the number of arrivals in a fixed interval, their
unordered times are independent uniform points in that interval.
In particular, if all arrivals in an interval of length $p$ are
served at its end, their expected total waiting is
\[
 \E[\text{total waiting}]=\lambda\int_0^p(p-t)\,dt
 =\frac{\lambda p^2}{2}.
\]
This is the basic calculation behind our periodic schedules.
The intervals in these statements are fixed independently of the
arrivals. Intervals selected by the offline optimum after seeing
requests require a separate argument; we do not assume the same
formulas apply to them. More general counting identities and the
potential arguments for optimal online policies are introduced only
in the technical development, Section~\ref{sec:technical-tools}.

\section{One resource: the value of knowing the future}\label{sec:scalar-story}
Before coordinating several resources, consider a single resource of
cost $w$ and requests arriving at rate $\lambda$. Every service buys
the resource once and may clear all pending requests. This is Poisson
TCP acknowledgment. We use it both to understand the best possible
online performance and to measure costs that remain unavoidable when
resources overlap.

\subsection{When is it worth collecting a batch?}
If purchases are cheap or requests are rare, immediate service is a
natural choice. When purchases are expensive and requests are frequent,
waiting for several requests can spread the purchase cost. The product
$q=\lambda w$ expresses this tradeoff. By measuring time in units of
$1/\lambda$ and multiplying all costs by $\lambda$, we obtain a
rate-one process with purchase cost $q$. Both cost rates and ratios
are preserved. We use these units throughout this section.

Consider serving every $n$th arrival, collecting $n$ requests per batch.
The mean time from one completed batch to the next is $n$. Each batch
costs $q$ to purchase. While collecting it, the queue successively has
$1,2,\ldots,n-1$ waiting requests, and each next arrival takes one unit
of time in expectation. Thus the mean waiting cost per batch is
$n(n-1)/2$. Dividing the mean total cost by the mean cycle length gives
\[
 \underbrace{\frac qn}_{\text{purchase cost rate}}
 +\underbrace{\frac{n-1}{2}}_{\text{waiting cost rate}}.
\]
Choosing the best integer $n$ therefore gives the candidate rate
\[
 g(q)=\min_{n\ge1}\left\{\frac qn+\frac{n-1}{2}\right\}.
\]
For $q<1$ the best choice is immediate service; for $1<q<3$ it is
two requests per batch; for $3<q<6$ it is three. At a boundary the
two neighboring choices tie. The algorithm uses the known load to
choose its batch size before requests arrive.

A count rule alone could leave the last, incomplete batch waiting
forever on a finite input. We add deadlines that grow with the batch
index. These deadlines always complete the input, while their total
expected extra cost is only logarithmic in the horizon. They do not
change the long-run rate. The explicit deadlines and their analysis
are given in Theorem~\ref{thm:r4-scalar-causal}.

\subsection{Why can no online policy do better?}
Optimizing over batch sizes does not yet exclude more elaborate
policies. An online algorithm could use the entire observed history,
randomize its decisions, or serve only part of the queue. To rule out
an improvement, we assign the pending queue a nonnegative amount of
unpaid cost, represented by a potential.

The potential is chosen so that waiting cost together with the expected
increase caused by new arrivals accumulates at rate at least $g(q)$.
A service can reduce this potential, but never by more than the
purchase cost it pays. Summing these two observations shows that every
online policy must pay at least $g(q)T$ in expectation. Completing the
last requests also pays off the final potential. This argument applies
to every causal policy, rather than only to count rules.

This is why the simple batching rule is enough: it attains a lower
bound that already accounts for every possible online decision.
Section~\ref{sec:r4-scalar} gives the potential explicitly and proves
both directions, including finite-input completion.

\subsection{What can the offline optimum save?}
The offline optimum can see whether the next request follows quickly
or only after a long silence. It can combine close arrivals and avoid
waiting through an unprofitable gap. To measure this advantage, let
$C(q)$ denote its long-run expected cost per unit time in the same
rate-one normalization.

Two elementary observations make this quantity accessible. First, an
optimal service can clear all pending requests and occur at the last
arrival in its batch. The problem is therefore to partition the
ordered arrivals into consecutive batches. Second, a gap of length
at least $q$ can separate two optimal batches: an extra purchase
costs $q$, while splitting saves at least that much waiting. We can
therefore divide the arrival sequence at these long gaps and optimize
each resulting cluster separately.

Poisson arrivals determine the distribution of cluster sizes and
internal gaps. A dynamic program computes the best partition inside
each finite cluster. Averaging these costs gives $C(q)$, with a
convergent expression and an explicit bound on the error from
discarding large clusters. The detailed construction is in
Section~\ref{sec:r2-single-exact}; rigorous numerical bounds are
given in Appendix~\ref{app:scalar-certificates}.

\begin{theorem}[Optimal Poisson TCP guarantee]\label{thm:tcp-overview}
For every load $q>0$, the optimal long-run ratio of expectations among
all causal completing policies is $g(q)/C(q)$. A deterministic policy
attains it. Over all positive loads,
\[
 \sup_{q>0}\frac{g(q)}{C(q)}=\frac1{C(1)}\approx1.409.
\]
The load $q=1$ attains this worst-case value.
\end{theorem}

At this hardest load, immediate service is already optimal online:
its expected cost rate is one. The offline rate is about $0.710$.
For comparison, at load $q=3$ the optimal ratio is below $4/3$, and
at $q=10$ it is below $1.25$. Thus the theorem describes the best
ratio at each load, not only one universal upper bound.

The remaining step in the worst-load proof is to compare different
loads. Concavity of $C(q)$ reduces this comparison to the loads where
the preferred batch size changes. A scaling inequality controls all
large such loads, and a few certified comparisons handle the small
ones. Corollary~\ref{cor:r4-scalar-ratio} and
Theorem~\ref{thm:r4-scalar-worst} give the complete proof.

This result also places a limit on general resource aggregation:
its instances include the one-resource example at load one. In the
other direction, the offline function $C$ will help us build a lower
bound for arbitrary overlapping requirements. It is this offline
comparison, rather than a separate TCP algorithm on each resource,
that we use next.

\section{Many resources: coordinating shared purchases}\label{sec:resource-story}
We now explain how one-resource comparisons lead to a constant
guarantee for arbitrary prescribed resource sets. The online algorithm
will use a periodic schedule. The central question is how to choose
its periods so that shared purchases remain inexpensive, and how to
compare this schedule with an offline optimum that can batch freely.

\begin{theorem}[General resource guarantee]\label{thm:resource-overview}
For independent Poisson arrivals of known rates, aggregation with
prescribed resources admits a deterministic online algorithm with
long-run ratio of expectations below $2.111$. For rational costs and
rates, a deterministic algorithm with polynomial-time preprocessing
achieves a ratio below $2.113$. Both guarantees are independent of
the number of resources, the number of request types, and their
pattern of overlap, and include completion of every finite input.
\end{theorem}

\subsection{Two difficulties created by overlap}
Suppose one type requires $\{a,b\}$ and another requires $\{b,c\}$.
Serving them together saves one purchase of $b$. If they use unrelated
service times, this saving may disappear. Conversely, making all
types wait for the same service time may impose too much delay on
frequent types. We therefore need different service frequencies that
still allow types to share purchases.

Overlap also makes lower bounds delicate. Solving an independent TCP
problem for each resource and adding its cost would charge the waiting
of a request once for every resource it requires. The original problem
charges that waiting only once. Such a sum need not be a valid lower
bound. The algorithm must coordinate purchases, and the analysis must
coordinate its charges to waiting cost.

\subsection{A lower bound that never charges the same cost twice}
We split costs into simpler problems, each involving one resource at
one time scale. A piece receives part of that resource's purchase
cost and part of the waiting cost of each type that requires it.
There are two basic budget rules:
\begin{itemize}
\item The purchase costs assigned to the pieces of a resource must
sum to at most its original purchase cost.
\item The waiting weights assigned to the pieces of a request must
sum to at most its original unit waiting rate.
\end{itemize}
For example, a request needing $\{a,b\}$ may contribute some of its
waiting cost to the $a$ pieces and the rest to the $b$ pieces. It does
not contribute its entire waiting cost to both.

Every original schedule induces a feasible schedule for each piece:
use the same service times for the relevant requests. The budget rules
ensure that the sum of the induced costs never exceeds the original
cost. Optimizing each piece separately can only reduce that sum.
Consequently, for every realized arrival sequence,
\[
 \OPT_T\ \ge\ \sum_{\text{pieces}}\OPT_T^{\text{piece}}.
\]
This comparison permits an unrestricted, clairvoyant original optimum.
The pieces are used only to measure what it must pay; its actual
services need not follow the decomposition.

A resource--scale LP chooses these budgets jointly. It also controls
the load in each piece so that the scalar offline analysis applies.
Independent thinning of Poisson arrivals converts the different
waiting weights into comparable one-resource instances. Averaging
the pathwise inequality then gives a lower bound proportional to
the LP optimum, denoted by $L^*$. The full LP and the projection
argument appear in Sections~\ref{sec:resource-lp}--\ref{sec:r2-sharp-lp}.

\subsection{Turning the same LP into a schedule}
The LP dual supplies a service time scale for each request type.
We round positive scales to periods of the form
\[
 h,\quad 2h,\quad 4h,\quad 8h,\quad\ldots
\]
with one common base $h$ and all clocks starting at zero. A type
assigned period zero is served immediately. A type with a positive
period is served at its clock ticks whenever its queue is nonempty.
If several types are due together, one service purchases the union
of their required resources.

The periods are nested: every tick of a longer period is also a tick
of each shorter period. For example, types with periods $2h$ and
$4h$ can share purchases at every multiple of $4h$, while the first
type still gets an additional opportunity halfway through. This is
how the schedule accommodates different frequencies without losing
all coordination. The periods depend only on the known instance,
and each actual purchase depends only on the requests already present.

This schedule also accommodates light demand. A zero time scale means
immediate service, with no rounding to a positive period. For a type
assigned a positive period, the algorithm waits only until its next
clock tick, not until another request of that type arrives. A sparse
type can therefore share purchases with frequent types without having
to wait for its own next arrival. The LP chooses these scales jointly,
taking both the arrival rates and the shared resource costs into account.

\paragraph{Waiting cost.}
A type with rate $\lambda_i$ and period $p_i$ incurs long-run expected
waiting cost $\lambda_i p_i/2$. This is the fixed-interval calculation
from Section~\ref{sec:model}: on average a request waits half a period.

\paragraph{Purchase cost.}
A clock tick does not always trigger a purchase. For one type of
rate $\lambda$ and period $p$, the probability that its queue is
nonempty is $1-e^{-\lambda p}$. The purchase cost rate for a resource
of cost $w$ used by that type alone is therefore
\[
 \frac{w}{p}\bigl(1-e^{-\lambda p}\bigr).
\]
With several types, we similarly count a shared resource only when
at least one due queue needs it. Nested periods and independent
Poisson streams let us calculate this probability. Retaining these
empty-queue savings is important for the constant, particularly for
infrequent types.

The dual constraints bound the combined waiting and purchase costs
by a constant multiple of the same $L^*$ used in the offline lower
bound. Choosing the common base carefully controls the loss in
rounding the periods. Averaging over possible choices proves that a
good one exists; deterministic preprocessing finds one before the
input arrives. No random choices are needed during execution.

\subsection{Where the constant comes from}
The proof has two estimates, both expressed through $L^*$.
With the selected parameters, their coefficients can be rounded
outward to give the convenient bounds
\[
 \E[\OPT_T]\ge0.6565\,L^*T,
 \qquad
 \E[\ALG_T]\le1.3856\,L^*T+O_I(1).
\]
Their ratio is below $2.111$. The important point is that the size
and overlap of the resource system enter through $L^*$ in both
inequalities. They cancel in the comparison, instead of appearing
as an extra depth, frequency, or resource-count factor. The lower
estimate uses the scalar offline cost; it does not multiply the
algorithm's cost by the worst TCP ratio.

The exact coefficients and their parameter choices yield the sharper
bound $2.110250205$ in Theorem~\ref{ag:pair:thm:main}. For a polynomial
implementation, we search a finite rational set of common bases and
certify the resulting costs to sufficient precision. This adds an
arbitrarily small multiplicative loss. Taking it to be at most
$1.001$ gives the stated bound below $2.113$; the bit-complexity
analysis is in Theorem~\ref{thm:efficient-resource}.

\subsection{Tree consequences and the remaining gap}
For rooted MLA, let each resource be an edge and each request type
require its path to the root. The union of these paths is precisely
the subtree needed to serve a batch. Both online and offline costs
are preserved, so the general theorem immediately improves the
$210$ bound reported for stochastic MLA, using an independent algorithm
and proof.

For prescribed pairs on a tree, a type requires the unique path
between its endpoints. A union of paths can have several connected
components. Serving those components simultaneously preserves the
total edge cost and every request's additive delay, so the theorem
also applies when each service must be connected, provided simultaneous
services carry no additional fee.

The TCP example gives a lower bound of about $1.409$ for all these
models. It does not settle whether arbitrary overlap is harder.
The general gap between this lower bound and our algorithm remains
open. The two main conclusions are thus different: a constant
algorithm for arbitrary resource requirements, and an optimal
characterization for the single-resource family.

\paragraph{Guide to the complete proofs.}
The argument above reduces the general guarantee to three tasks:
justify the one-resource projections, bound the cost of the nested
schedule by the same LP, and implement its preprocessing efficiently.
Section~\ref{ag:sec:pair-additive} proves these in order.
Section~\ref{sec:scalar} proves the scalar cost formulas and the
worst-load comparison. The probability and potential tools needed
by these proofs are collected in Section~\ref{sec:technical-tools}.
Appendix~\ref{app:scalar-certificates} gives the numerical certificates.

\section{Technical tools for the proofs}\label{sec:technical-tools}
Sections~\ref{sec:scalar-story} and~\ref{sec:resource-story} gave the
algorithms, guarantees, and their main comparisons. We now develop
the complete proofs. This section records the probability identities
and queue accounting used below; the next two sections prove the
TCP and resource guarantees, respectively.

The first subsection supplies Poisson counting identities, including
the insertion formula used to count scalar clusters. The second
formalizes the unpaid-cost argument and the completion rule from
Section~\ref{sec:scalar-story}. Neither queue lemma is needed to
understand or execute the prescribed-resource algorithm.

We use $z_+=\max\{z,0\}$ for the positive part, $\mathbf1_A$ for an
event indicator, and $|J|$ for the length of a measurable time set.
Charging blocks are half-open: an action at the left endpoint belongs
to the block and one at its right endpoint belongs to the next.
Nonnegative terms permit interchanging sums, integrals, and
expectations by Tonelli's theorem.

\subsection{Poisson counting identities}\label{sec:poisson}
For a rate-$\lambda$ process, the number $N(J)$ of arrivals in an
interval $J$ has distribution
$$
 \Pr(N(J)=k)=e^{-\lambda|J|}\frac{(\lambda|J|)^k}{k!},
 \qquad \E[N(J)]=\operatorname{Var}(N(J))=\lambda|J|.
$$
Counts in disjoint intervals are independent. Conditional on $n$
arrivals in a finite interval, their unordered times are $n$
independent uniform points in that interval. In particular, for a
nonnegative deterministic function $g$,
\begin{equation}\label{eq:poisson-sum}
 \E\left[\sum_{a_i\in[0,T]}g(a_i)\right]
   =\lambda\int_0^T g(t)\,dt.
\end{equation}
Conditioning on the count proves this identity: the conditional
expectation is $nT^{-1}\int_0^Tg(t)\,dt$, and $\E[n]=\lambda T$.
We use it, for example, with $g(t)$ equal to the wait until the next
deterministic clock tick. Nonnegative integrands also allow us to
exchange sums, integrals, and expectations by Tonelli's theorem.

For independent streams, superposition adds their rates. Conversely,
independently assigning an arrival a mark $j$ with probability $q_j$
produces independent Poisson streams of rates $q_j\lambda$. The
generating function
$\E[\prod_j z_j^{N_j}]=\exp(\lambda|J|\sum_jq_j(z_j-1))$
factors into the generating functions of these independent counts.
These statements apply to marks chosen independently for each
occurrence; arbitrary routing based on the future need not preserve
a Poisson law.

The void probability is $\Pr(N(J)=0)=e^{-\lambda|J|}$.
It remains valid for a deterministic interval that depends on the
instance parameters. We will not apply it to intervals selected by
an offline schedule after observing arrivals. Instead, our lower
bounds first establish an inequality for every realized schedule
and then average a certificate whose counting regions are fixed
independently of that schedule.

We also use the following extension of~\eqref{eq:poisson-sum}, in
which the contribution of a distinguished arrival depends on the
other arrivals. It will count scalar cluster starts without conditioning on an arrival
at one prescribed time.

\begin{lemma}[Poisson insertion]\label{lem:poisson-insertion}
Fix $x$, and let $u_x(a,J)$ be a nonnegative measurable quantity defined
for a distinguished type-$x$ request at time $a$ and a finite multiset $J$
of other requests. Then
\begin{equation}\label{eq:poisson-insertion}
 \E\!\left[\sum_{i\in I_T:\,x_i=x}
   u_x(a_i,I_T\setminus\{i\})\right]
 =\lambda_x\int_0^T\,\E[u_x(a,I_T)]\,da.
\end{equation}
On the right, $I_T$ is an independent Poisson input to which the
distinguished request can be added inside the definition of $u_x$.
\end{lemma}
\begin{proof}
Condition on the number $N_x=n$ of type-$x$ arrivals. Their times are
independent uniforms in $[0,T]$, and exchangeability makes the sum on
the left equal in expectation to $n$ times the contribution of one
distinguished uniform time, with $n-1$ other type-$x$ times. The identity
$n\Pr(N_x=n)=\lambda_xT\Pr(N_x=n-1)$ changes the latter $n-1$ points
back into an unconditional type-$x$ Poisson process. All other types
retain their independent laws. Integrating the distinguished uniform
time proves~\eqref{eq:poisson-insertion}. No conditioning on an arrival at one
prescribed time is needed.
\end{proof}

\subsection{Potential accounting and finite completion}\label{sec:potential-tools}
Several proofs use the same queue calculation. Write $Q(t)$ for the
pending-count vector, $H(q)$ for its waiting rate, and
$\mathcal LV(q)=\sum_i\lambda_i[V(q+e_i)-V(q)]$ for the arrival
generator. At a service $q\to q'$ of cost $c$, call
$c+V(q')-V(q)$ its excess. Arrivals are inserted before their triggered
services, as in our event convention.

\begin{lemma}\label{lem:potential-accounting}
Let $V$ be bounded and nonnegative, with $V(0)=0$.
For a causal policy starting empty,
$$
 \E[\cost_{[0,T]}+V(Q(T))]
 =\E\left[\int_0^T(H+\mathcal LV)(Q(t))\,dt\right]
   +\E\left[\sum_{\text{services through }T}[c+V(q')-V(q)]\right].
$$
Here $\cost_{[0,T]}$ includes only costs incurred through $T$.
If $H+\mathcal LV\ge g$ in every state and every service has
nonnegative excess, then every causal completing policy satisfies
$\E[\ALG_T]\ge gT$, including randomized policies and partial services.
\end{lemma}
\begin{proof}
Telescope the potential over arrivals and services. Conditional
Poisson compensation replaces the arrival increments by
$\int_0^T\mathcal LV(Q(t))dt$; independent policy coins are included
in the history. Boundedness permits localization for policies of
finite expected cost. The lower-bound hypotheses give
$\E[\cost_{[0,T]}+V(Q(T))]\ge gT$. After the last arrival, the
service inequalities telescope from $V(Q(T))$ to $V(0)=0$, so
completion pays this terminal potential. Infinite expected cost
already satisfies the lower bound.
\end{proof}

The next lemma adds finite deadlines to a count-based policy without
changing its asymptotic rate. A busy period starts when its queue
becomes nonempty and ends when it becomes empty. Normal services
occur only at arrivals.

\begin{lemma}\label{lem:timeout-completion}
With $V$ as above and unit waiting per request, suppose an untimed
policy has at most $n$ pending requests between
events, $(H+\mathcal LV)(q)\le g$ in every state it occupies between
events, including the empty state, and
nonpositive excess at every normal service. Assume clearing any
occupied state has excess at most $b\ge0$. Conditional on the past
at each busy-period start, its untimed duration $S$ satisfies
$\Pr(S>t\mid\mathrm{past})\le Be^{-\eta t}$ for $t\ge0$,
where $\eta>0$ and $B\ge1$.
Suppose the expected number of starts through $T$ is at most $\nu T$.
Clear period $m$ at its normal end or at the earlier deadline
$$
 H_m=\frac{\ln B+2\ln(m+1)}{\eta}.
$$
Then every finite input completes without an end signal, and
$$
 \E[\ALG_T]\le gT+2b+
 \frac n\eta\bigl[\ln B+2\ln(\nu T+1)\bigr].
$$
\end{lemma}
\begin{proof}
The tail bound makes the conditional probability of timeout
$m$ at most $(m+1)^{-2}$. Their sum is less than one. The accounting
identity gives $\E[\cost_{[0,T]}+V(Q(T))]\le gT+b$ on an infinite
continuation. The truncated input has the same prefix. If period $m$
is pending at $T$, it clears by its deadline, paying at most
$V(Q(T))+b+nH_m$ afterward. Its index is at most the number of
starts through $T$. Jensen's inequality proves the bound.
Every nonempty period has a finite deadline. The tail bound must hold
after every history of the timed policy; the start-count bound refers
to that policy as well.
\end{proof}

\section{Proof of the optimal Poisson TCP guarantee}\label{sec:scalar}
With one resource, the model is TCP acknowledgment with additive delay:
one purchase clears pending requests, and every request pays its own
waiting time. We study its known-rate Poisson arrival version.
The scalar problem has two roles. Its offline rate gives the resource
LP certificate, while the difference between its optimal causal and
clairvoyant rates gives a lower bound for every online algorithm.
Neither role restricts the offline batch sizes.

For one resource of cost $w$ and arrival rate $\lambda$, put
$q=\lambda w$. The time change $t\mapsto\lambda t$, together with
multiplication of cost by $\lambda$, gives rate one and setup cost
$q$. Cost per unit time and ratios of expectations are unchanged.
The formulas below use these normalized units. In particular, an
additive finite-horizon constant must be divided by $\lambda$ when
returning to the original units.
\subsection{The exact unrestricted offline rate}
\label{sec:r2-single-exact}

For $n$ arrivals at times $a_1=0<a_2<\cdots<a_n$, write
$x_j=a_{j+1}-a_j$.
Define $D_0=0$ and, for $1\le j\le n$,
$$
 D_j(q;x_1,\ldots,x_{j-1})
 =\min_{0\le k<j}\left\{
 D_k+q+\sum_{r=k+1}^{j-1}(r-k)x_r\right\}.
$$
An empty sum is zero.
Thus $D_1=q$, and the cost of serving three arrivals in one batch is
$q+x_1+2x_2$.
The factor on a gap is the number of requests already waiting while
that gap elapses.

\begin{lemma}
The dynamic program $D_n$ gives the unrestricted offline optimum for
these $n$ arrivals.
Moreover, any gap of length at least $q$ can be made a boundary between
two batches in an optimum.
\end{lemma}
\begin{proof}
At any service, clearing all pending requests costs no more than clearing
a subset, and weakly decreases all affected waiting costs.
Deleting any resulting empty later services yields a schedule whose
batches are consecutive segments of the arrival sequence.
Each service can then be moved to the last arrival in its batch, which
weakly decreases waiting and preserves the order of services.
The final batch after index $k$ has exactly the cost displayed in the
recurrence, proving completeness of the dynamic program.

Suppose one such batch spans a gap $x_j\ge q$.
Split it at that gap and serve its earlier part at the last arrival of
that part.
The extra purchase costs $q$, whereas at least one earlier request
avoids waiting for the entire gap, saving at least $q$.
The split therefore cannot increase cost.
Repeating the operation establishes the second assertion.
\end{proof}

Put $p=e^{-q}$ and $r=1-p$.
Consecutive arrivals separated by gaps smaller than $q$ form a cluster;
a gap at least $q$ separates clusters.
For a cluster beginning at a specified arrival, its number of arrivals
$N$ has distribution
$$
 \Pr(N=n)=p r^{n-1},\qquad n\ge1.
$$
Conditional on $N=n$, the internal gaps are independent exponential
variables conditioned to belong to $[0,q)$.
Their joint density is
$e^{-\sum_jx_j}/r^{n-1}$ on that cube.
The exact cost rate is described by the following absolutely convergent
series of finite-dimensional integrals.

\begin{theorem}
For every $q>0$, define
$$
 C(q)=p^2\sum_{n=1}^{\infty}
 \int_{[0,q]^{n-1}}
 D_n(q;x_1,\ldots,x_{n-1})
 e^{-\sum_{j=1}^{n-1}x_j}\,dx_1\cdots dx_{n-1}.
$$
For $n=1$, the integral denotes the number $q$.
Then the unrestricted offline optimum satisfies
$$
 \left|\mathbb E[\operatorname{OPT}_T]-C(q)T\right|
 \le4qe^q
 \qquad(T\ge0).
$$
In particular its expected long-run cost rate exists and equals $C(q)$.
If $C_K(q)$ denotes the series truncated at $n=K$, then
$$
 0\le C(q)-C_K(q)
 \le q r^K(1+Kp).
$$
Also $q p^2\le C(q)\le q$.
\end{theorem}
\begin{proof}
Extend the rate-one Poisson input to the entire real line.
The preceding lemma shows that optimal costs add over its clusters.
A distinguished arrival begins a cluster precisely when there is no
arrival in the preceding interval of length $q$.
This event has probability $p$ and is independent of the entire future
cluster.
Thus cluster starts have intensity $p$.
If $D$ denotes the cost of a whole cluster beginning at a distinguished
arrival, the Poisson insertion identity gives
$$
 \mathbb E\left[\sum_{\substack{\text{cluster starts }s\in[0,T]}}
 D(\text{cluster at }s)\right]=Tp\,\mathbb E[D].
$$
This application can be reduced to the finite-window
insertion identity: truncate each cluster after its first $K$ arrivals.
The resulting quantity depends only on the Poisson configuration in
$[-q,T+Kq]$.
Let $K\to\infty$.
Deleting requests cannot increase optimum cost, so the truncated costs
increase to $D$, and monotone convergence applies.
Also $D\le qN$ and $\mathbb E[N]=1/p$, ensuring integrability.

Conditioning on $N$ and its internal gaps gives
$$
 p\,\mathbb E[D]
 =p^2\sum_{n\ge1}\int_{[0,q]^{n-1}}D_n e^{-\sum_jx_j}\,dx
 =C(q).
$$
All clusters wholly inside $[0,T]$ contribute exactly the same amount
to this stationary reward sum and to $\operatorname{OPT}_T$.
Only clusters crossing $0$ or $T$ can contribute a difference.
Deleting arrivals from a cluster cannot increase its optimal cost, and
its full cost is at most $q$ times its full size.
At any deterministic boundary, the size of a cluster crossing it is
bounded by the sum of two geometric$(p)$ counts: scan backward from
the nearest arrival on the left and forward from the nearest arrival
on the right, stopping each scan at the first gap at least $q$.
Each count has expectation $1/p$.
If the gap straddling the boundary is itself at least $q$, there is no
crossing cluster and its contribution is zero.
Hence the expected size charged at a boundary is at most $2/p$.
The two boundaries together yield
$$
 \left|\mathbb E[\operatorname{OPT}_T]-Tp\,\mathbb E[D]\right|
 \le\frac{4q}{p}=4qe^q.
$$
The same estimate covers the case where one cluster crosses both
boundaries, since counting it twice only enlarges the bound.

Finally $q\le D_n\le nq$ for every nonempty cluster.
In particular the first series term gives $C(q)\ge qp^2>0$.
The upper bounds imply $C(q)\le q$ and
$$
 \begin{aligned}
 C(q)-C_K(q)
 &\le p^2q\sum_{n=K+1}^{\infty}n r^{n-1}\\
 &=q r^K\bigl((K+1)p+r\bigr)
 =q r^K(1+Kp).
 \end{aligned}
$$
This proves convergence and the explicit tail error.
\end{proof}

\subsection{The exact causal optimum for every scalar load}
\label{sec:r4-scalar}

Continue in the rate-one normalization, with setup cost $q$.

Define
$$
 g(q)=\min_{n\ge1}\left\{\frac qn+\frac{n-1}{2}\right\}.
$$
Choose an integer $N\ge1$ with
$$
 \frac{N(N-1)}2\le q\le\frac{N(N+1)}2.
$$
Then $g(q)=q/N+(N-1)/2$ and $N-1\le g(q)\le N$.
At a common endpoint either adjacent choice is valid.
These facts follow by comparing consecutive terms: their difference is
$1/2-q/[n(n+1)]$.

\begin{theorem}\label{thm:r4-scalar-causal}
For every $q>0$, every causal completing policy, including randomized
policies and policies using partial batches, satisfies
$$
 \mathbb E[\operatorname{ALG}_T]\ge g(q)T\qquad(T\ge0).
$$
There is a deterministic horizon-unaware policy that completes every
finite input and has expected cost rate exactly $g(q)$.
For $N\ge2$, it satisfies the explicit bound
$$
 \mathbb E[\operatorname{ALG}_T]
 \le g(q)T+2q+(N-1)\bigl[2(N-1)\ln2+4\ln(T+1)\bigr].
$$
For $N=1$, immediate service has expected cost exactly $qT$.
\end{theorem}

\begin{proof}
Write $g=g(q)$ and define
$$
 h_q(n)=
 \begin{cases}
 ng-n(n-1)/2,&0\le n\le N,\\
 q,&n\ge N.
 \end{cases}
$$
The formulas agree at $N$. Since $N-1\le g\le N$, the potential
is nondecreasing and lies in $[0,q]$. Moreover,
$$
 n+h_q(n+1)-h_q(n)\ge g,
$$
with equality for $n<N$. Every service costs $q$ and lowers $h_q$
by at most $q$, including a partial clearing.
Lemma~\ref{lem:potential-accounting} proves the all-causal lower bound.

For the upper bound, serve immediately when $N=1$.
Otherwise clear a batch at its $N$th arrival or at the deadline
$$
 H_m=2(N-1)\ln2+4\ln(m+1)
$$
after the first arrival of its $m$th batch. Between events the queue
has at most $N-1$ requests and drift plus holding exactly $g$.
A full batch costs its potential decrease $h_q(N)=q$; a timeout
has excess at most $q$. The untimed wait for $N-1$ new arrivals
satisfies $\E[e^{S/2}]=2^{N-1}$, hence
$\Pr(S>t)\le2^{N-1}e^{-t/2}$. At most one batch starts per arrival.
Lemma~\ref{lem:timeout-completion}, with
$(n,b,\eta,B,\nu)=(N-1,q,1/2,2^{N-1},1)$, gives precisely the
stated finite-horizon bound. Its excess is $O(\ln(T+1))$, proving
attainment of $g$ with full completion.
\end{proof}

\begin{corollary}\label{cor:r4-scalar-ratio}
For every $q>0$, the smallest long-run ratio of expectations among
causal completing policies is exactly $g(q)/C(q)$.
One deterministic policy, using the known instance parameters to
select the rule above, attains this value on every scalar instance.
\end{corollary}
\begin{proof}
The offline expected cost rate exists and is $C(q)>0$.
Divide the finite-horizon lower bound and the attaining policy's
asymptotic equality by $\mathbb E[\operatorname{OPT}_T]$.
\end{proof}

\subsubsection{Two structural properties of the scalar offline rate}

\begin{lemma}\label{lem:r4-scalar-concavity}
The function $C$ is concave on $[0,\infty)$, with $C(0)=0$.
Writing $q_n=n(n+1)/2$, one has
$$
 \sup_{q>0}\frac{g(q)}{C(q)}
 =\sup_{n\ge1}\frac{n}{C(q_n)}.
$$
\end{lemma}
\begin{proof}
For each fixed arrival realization, the offline cost is the infimum
over schedules of $q$ times the number of services plus their waiting
cost.
It is therefore concave in $q$.
Taking expectations, dividing by the horizon, and passing to the
existing cost-rate limit preserves concavity.
The free-setup problem has zero cost, and $0\le C(q)\le q$, so the
extension at zero is valid.

On $[q_{n-1},q_n]$, the function $g$ is affine with endpoint values
$n-1,n$.
Concavity lower-bounds $C$ by the chord between the same endpoints.
It follows that $C(q)/g(q)$ is at least the smaller endpoint ratio.
For the first interval $(0,1]$, use $C(q)\ge qC(1)$ and $g(q)=q$
instead of forming the undefined ratio at zero.
Every positive $q$ belongs to one of these intervals, proving the
supremum identity.
\end{proof}

\begin{lemma}\label{lem:r4-sqrt-monotonicity}
For $s\ge1$ and $q>0$,
$$
 C(s^2q)\ge sC(q).
$$
Equivalently $C(q)/\sqrt q$ is nondecreasing.
\end{lemma}
\begin{proof}
Let $J$ be a finite realized set of arrival slots, and define
$$
 F_J(z)=\inf_{S\text{ completes every slot of }J}
 \left\{q\,\#\mathrm{services}(S)+\sum_{i\in J}z_iW_i(S)\right\},
 \qquad z_i\ge0.
$$
The feasible family is fixed, including when some $z_i$ are zero.
Thus $F_J$ is concave as an infimum of affine functions of $z$.
Independently retain each slot with probability $1/s$, writing $B_i$
for its retention indicator and $J_B$ for the retained input.
Conditional Jensen gives
$$
 F_J(\mathbf1/s)\ge\,\mathbb E_B[F_J(B)]
 \ge\,\mathbb E_B[\operatorname{OPT}_{q}(J_B)].
$$
The second inequality follows by deleting the zero-weight slots from
any schedule completing all slots.
Deleting them can only remove feasibility constraints and can never
increase the cost of serving the retained requests.
In particular we do not move a zero-weight request to a time preceding
its arrival, nor claim equality of these two optima.

Now let $J$ be a rate-$s$ Poisson input in $[0,T]$.
Its independently retained subset has rate one.
Factoring $1/s$ out of the objective gives
$F_J(\mathbf1/s)=s^{-1}\operatorname{OPT}_{sq}(J)$.
Consequently, for every finite horizon including completion,
$$
 \frac1s\,\mathbb E[\operatorname{OPT}_{\lambda=s,w=sq,T}]
 \ge\,\mathbb E[\operatorname{OPT}_{\lambda=1,w=q,T}].
$$
The normalized scalar cost rate for parameters $(\lambda,w)$ is
$C(\lambda w)$.
Divide by $T$ and pass to the existing limits to obtain the assertion.
\end{proof}

\subsubsection{The worst scalar load is exactly one}

Appendix~\ref{app:scalar-certificates} certifies the following four strict rational inequalities:
$$
 C(1)<\frac57,\qquad
 C(3)>\frac32,\qquad
 C(6)>\frac94,\qquad
 C(10)>\frac{16}{5}.
$$
\begin{theorem}\label{thm:r4-scalar-worst}
The optimal competitive factor over all positive-rate, positive-setup
single-resource instances is exactly
$$
 \boxed{\ \sup_{q>0}\frac{g(q)}{C(q)}=\frac1{C(1)}.\ }
$$
The load $q=1$ attains the worst factor.
This is a matching upper and lower characterization for the entire
scalar family, including unrestricted clairvoyant batching and all
causal randomized policies.
\end{theorem}
\begin{proof}
For the second and third triangular loads,
$$
 C(3)>\frac32>2\frac57>2C(1),\qquad
 C(6)>\frac94>3\frac57>3C(1).
$$
Furthermore
$$
 \left(\frac{16}{5}\right)^2
 -20\left(\frac57\right)^2=\frac{44}{1225}>0,
$$
so $C(10)>\sqrt{20}\,C(1)$.
For every $n\ge4$, the triangular load $q_n\ge10$ satisfies, by
Lemma~\ref{lem:r4-sqrt-monotonicity},
$$
 C(q_n)\ge\sqrt{q_n/10}\,C(10)
 >\sqrt{2q_n}\,C(1)
 =\sqrt{n(n+1)}\,C(1)>nC(1).
$$
Thus every triangular load after the first has a strictly smaller
ratio than $1/C(1)$.
Lemma~\ref{lem:r4-scalar-concavity} controls every intervening load.
At $q=1$, $g(1)=1$, so equality holds.
Corollary~\ref{cor:r4-scalar-ratio} supplies both the universal causal
lower bound and a completing policy attaining the upper bound.
\end{proof}

\subsection{A numerical lower bound for resource aggregation}
The finite-state and conditional-mean certificates in
Appendix~\ref{app:scalar-certificates} give
$$
 0.709707<C(1)<0.710122517893,
 \qquad 1.4082077033<\frac1{C(1)}<1.40904.
$$
The lower bound on $C(1)$ controls the upper end of the scalar
factor; the upper bound on $C(1)$ supplies the competitive lower
bound. The directions are different and both are needed.

\begin{theorem}[Competitive lower bound]\label{thm:general-lower}
Every causal algorithm for prescribed-resource aggregation, including
randomized algorithms, has worst-instance long-run ratio at least
$1/C(1)>1.4082077033$. The same lower bound holds for rooted and prescribed-pair
aggregation on trees.
\end{theorem}
\begin{proof}
Take one resource of cost one with a rate-one stream. Every causal
completing policy costs at least $T$ in expectation by
Theorem~\ref{thm:r4-scalar-causal}, while the offline cost rate is
$C(1)$. A single unit edge realizes the instance for rooted and pair
aggregation. These embeddings preserve all waiting, purchases,
and completion costs.
\end{proof}
Thus the optimal uniform resource factor $R^*_{\rm resource}$ obeys
$$
 1.4082077033<R^*_{\rm resource}<2.110250205.
$$
The exact scalar theorem supplies a matching result only within the
single-resource family. Resource overlap may still affect the optimal
uniform factor.

\section{Proof of the prescribed-resource guarantee}\label{ag:sec:pair-additive}

Each type requires a fixed resource set, and these sets may overlap
arbitrarily. A finite LP chooses one time scale per type. After one
common phase is selected, the online rule is simply to clear the due
nonempty queues and buy their resource union. We prove a ratio below
$2.110250205$ against the unrestricted offline optimum.

\subsection{The problem and its tree interpretations}
\paragraph{Notation.}
We use the model of Section~\ref{sec:model}: resource $e\in E$
has purchase cost $w_e$, and type $i\in I$ requires $R_i\subseteq E$
and arrives at rate $\lambda_i$. Remove unused and zero-cost resources,
zero-rate types, and types whose remaining requirements are empty;
the latter are served immediately for free. If nothing remains the
problem is trivial. Otherwise costs and rates are positive and all
required sets are nonempty. Each occurrence receives its entire
required set in one service, which pays the weight of the union;
purchases are repeated in later services.

In a tree, $R_i$ is the unique path between the two endpoints of type $i$.
Buying the union of these paths gives a forest. Each component may be
purchased as a separate connected service at the same time, preserving
both total edge cost and each request's additive delay. Thus the theorem
also applies when each individual action must buy a connected subtree,
provided simultaneous actions are permitted without additional fees.
The line case is the union-of-intervals specialization.

For rooted aggregation, take one type at each positive-rate client
vertex $x$ and set $R_x$ to its path to the prescribed root. The union
of the paths of a nonempty batch is a connected rooted subtree, and
trimming any rooted service to this union cannot increase its cost.
Thus the resource formulation has exactly the same optimum. A fixed
root activation can be represented by one additional resource required
by every type. Zero-cost resources may be contracted or removed;
requests whose remaining required set is empty are served for free.

The one-resource tradeoff was developed in
Section~\ref{sec:scalar-story}. With several resources, independent
clocks can repeatedly purchase a shared resource that one offline
batch buys only once. We therefore allocate purchase and waiting
budgets jointly before choosing the service periods.

\subsection{A lower-bound LP with no repeated charging}
\label{sec:resource-lp}
Let $\mathcal A$ be a finite set of positive time scales and fix
$\beta>0$. For each resource--scale pair use a purchase budget
$c_{ea}$, and for each incidence $e\in R_i$ use a charge
$\delta_{iea}$. The program has objective
$$
 L^*=\max\sum_{i,e\in R_i,a\in\mathcal A}\lambda_i\delta_{iea}
$$
and nonnegative variables satisfying
$$
 \sum_{e\in R_i,a}\frac{\delta_{iea}}a\le1,\qquad
 \delta_{iea}\le c_{ea},\qquad
 a\sum_{i:e\in R_i}\lambda_i\delta_{iea}\le\beta c_{ea},\qquad
 \sum_a c_{ea}\le w_e.
$$
The first and last inequalities allocate the original waiting and
purchase budgets. The other two control the size and mean load of
the weighted scalar problem at a resource--scale pair. The zero
solution is feasible, and all variables are bounded, so an optimum
exists. We will prove the paired estimates
$$
 \E[\OPT_T]\ge\frac{C(\beta)}\beta L^*T,
 \qquad
 \E[\ALG_T]\le\left(r+\frac{D}{\beta}\right)L^*T+K_I,
 \qquad D=\frac1{4(\ln2)^2}.
$$
Here $r>1$ will be the finite grid ratio, and $K_I$ includes the final
clearings. We first establish the lower certificate and then derive
the clocks from the dual.

\subsection{The scalar projection certificate}
\label{sec:r2-sharp-lp}

For a rate-one Poisson process, let $F_q(T)$ be the expected unrestricted
offline batching cost on $[0,T]$, with purchase cost $q>0$ and unit
waiting cost. Define
$$
 C(q)=\lim_{T\to\infty}\frac{F_q(T)}T.
$$
The exact convergent series and existence of this limit are proved in
Section~\ref{sec:r2-single-exact}. In particular $0<C(q)\le q$.

\begin{lemma}\label{lem:r2-scalar-monotonicity}
For every $q,T>0$, $F_q(T)\ge C(q)T$.
The ratio $C(q)/q$ is nonincreasing in $q$.
For a rate-$u$ Poisson process with unit purchase cost and unit waiting
cost, its expected offline cost on $[0,H]$ is at least $C(u)H$.
\end{lemma}
\begin{proof}
Every finite single-resource input has an optimum that completes at
its last arrival, as proved with the batching dynamic program below.
Concatenating optimal schedules on two adjacent intervals is feasible.
Independent stationary Poisson increments therefore give
$F_q(S+T)\le F_q(S)+F_q(T)$.
For any positive integer $n$, $F_q(nT)\le nF_q(T)$.
Divide by $nT$ and use the defining limit to obtain the first claim.

For each fixed finite input, its optimum divided by $q$ equals
$$
 \inf_{\mathcal S}\left\{\text{number of purchases in }\mathcal S
                +\frac{\text{total waiting in }\mathcal S}{q}\right\}.
$$
This quantity is nonincreasing in $q$.
Taking expectations and then the long-run limit proves the second claim.
Finally the time change $t'=ut$, together with multiplying cost by $u$,
converts rate $u$ and purchase cost $1$ into rate $1$ and purchase cost
$u$. Its expected cost is $F_u(uH)/u\ge C(u)H$.
\end{proof}

Let $L$ be the objective value of any feasible solution of the program above.

\begin{theorem}[Sharp coefficient for the LP certificate]
\label{thm:r2-sharp-lp}
For every feasible LP solution and every $T\ge0$,
$$
 \mathbb E[\operatorname{OPT}_T]\ge\frac{C(\beta)}\beta LT.
$$
The coefficient $C(\beta)/\beta$ cannot be increased uniformly over
the prescribed-resource instances and finite grids defined below.
This is a sharpness assertion about the LP certificate, not about the
competitive ratio of an online algorithm.
\end{theorem}
\begin{proof}
Fix a realized finite input. For each resource--scale pair $(e,a)$,
form a separate optimization problem with purchase cost $c_{ea}$
and, for each original type-$i$ occurrence requiring $e$, waiting
weight $\delta_{iea}/a$. Zero-weight occurrences may be removed.
Let its unrestricted optimum be $O_{ea}$.
Any feasible original schedule induces a feasible schedule for each
of these subproblems: serve an occurrence when the original schedule
serves it and use the corresponding purchase of $e$.
Sum the subproblem costs. The budget $\sum_a c_{ea}\le w_e$ pays
for each resource purchase, and the budget
$\sum_{e\in R_i,a}\delta_{iea}/a\le1$ pays for each request's waiting.
Taking the infimum over original schedules yields the pathwise bound
$$
 \operatorname{OPT}_T\ge\sum_{e,a}O_{ea}.
$$

Fix a pair with $c=c_{ea}>0$.
Independently retain each type-$i$ occurrence with probability
$\delta_{iea}/c$ and give every retained occurrence waiting weight
$c/a$. For any fixed schedule of the original weighted subproblem,
the expected retained waiting is exactly its original weighted waiting.
Apply its purchases to the retained input; purchases clearing no
retained request can be deleted. The resulting expected cost is no
larger than the original weighted schedule cost and is at least the
expected retained optimum. Since the latter is independent of the
chosen schedule, taking its infimum gives
$$
 O_{ea}\ge\,\mathbb E_{\rm thinning}[O^{\rm retained}_{ea}].
$$
This argument holds after fixing the input, even though the minimizing
schedule is clairvoyant.

Now average the input. Independent Poisson thinning and superposition
make the retained process homogeneous. Normalize time by $a$ and cost
by $c$. The result is a unit-purchase, unit-waiting problem of rate
$$
 u=\frac a c\sum_i\lambda_i\delta_{iea}\le\beta
$$
and horizon $H=T/a$.
Lemma~\ref{lem:r2-scalar-monotonicity} and its monotonicity assertion give
$$
 \mathbb E[O_{ea}]\ge c C(u)\frac T a
 \ge\frac{C(\beta)}\beta\,T\sum_i\lambda_i\delta_{iea}.
$$
If $c=0$ or $u=0$, the desired contribution is zero.
Summing proves the full finite-horizon lower bound without a boundary
loss or a restriction on the offline comparator.

For sharpness take one resource of cost $1$ and one type of rate
$\lambda=\beta$. The finite grid below contains $a=1$:
its lower endpoint is at most $1/2$, and its upper endpoint is at least
$2$. The feasible choice $\delta_{1,e,1}=c_{e,1}=1$, zero elsewhere,
has $L=\beta$; immediate service also supplies a dual solution of value
$\beta$, so $L^*=\beta$.
Time scaling makes the offline asymptotic cost rate exactly $C(\beta)$.
Thus a larger uniform coefficient would fail as $T\to\infty$ on this
single-resource instance.
\end{proof}

\subsection{Using the exact probability of an empty clock tick}
\label{sec:r2-empty-ticks}

We now choose the scale grid and use the complete prefix information in the dual. Empty scheduled clearings incur no purchase cost.

Write $\ell=\ln2$ and $D=1/(4\ell^2)$.
Fix a grid ratio $r>1$ and a mean-budget parameter $\beta>0$, and set
$$
 K=r+D/\beta,\qquad s=2\ell K.
$$
Form a finite LP grid from consecutive powers of $r$.
Its smallest scale is the largest such power at most $\beta/(2\Gamma)$, and its largest scale is the smallest such power at least
$$
 \max\{2M/\lambda_{\min},\beta/\lambda_{\min}\}.
$$
Here $\Gamma=\sum_i\lambda_i$, $\lambda_{\min}=\min_i\lambda_i$, $C_i=\sum_{e\in R_i}w_e$, and $M=\sum_i\lambda_i C_i$.
The dual minimizes $\sum_i y_i+\sum_e w_ez_e$ over nonnegative
variables subject to
$$
 y_i/a+s_{iea}+a\lambda_i t_{ea}\ge\lambda_i,\qquad
 z_e\ge\beta t_{ea}+\sum_{i:e\in R_i}s_{iea}.
$$
Solve these finite resource primal and dual programs on this grid and let their common optimum be $L^*$.
For an optimal dual put $\tau_i=y_i/\lambda_i$.
Setting $y_i=0$, $t_{ea}=0$, $s_{iea}=\lambda_i$, and $z_e=\Lambda_e$ gives a feasible dual of value $M$. Strong duality therefore implies
$$
 L^*\le M,\qquad 0\le\tau_i\le M/\lambda_{\min}.
$$

\subsubsection{The cumulative information in the dual}

For a resource $e$, let $\Lambda_e=\sum_{i:e\in R_i}\lambda_i$.
Sort its types by $\tau_i$ and regard each type as an interval of length $\lambda_i$ carrying the constant value $\tau_i$.
This defines a nondecreasing rate quantile $\tau_e(v)$ for $0<v<\Lambda_e$.
Put
$$
 G_e(v)=\int_0^v\tau_e(u)\,du,\qquad
 Z_e=\max_{0\le v\le\Lambda_e}v(1-G_e(v)/\beta).
$$
The function $G_e$ is continuous, nonnegative, nondecreasing, convex, and piecewise linear.
Both $G_e$ and $Z_e$ are computable by sorting finitely many values and maximizing finitely many quadratic polynomials.

\begin{lemma}\label{lem:r2-empty-prefix}
For every used resource, $0<Z_e<r z_e$.
\end{lemma}
\begin{proof}
Fix a grid scale $a$ with $v=\beta/a\le\Lambda_e$.
Taking a fractional prefix of the sorted types gives numbers $\alpha_i\in[0,1]$ with
$$
 \sum_i\alpha_i\lambda_i=v,\qquad
 \sum_i\alpha_i\lambda_i\tau_i=G_e(v).
$$
The type dual constraints and nonnegativity imply
$$
 \sum_i s_{iea}\ge\sum_i\alpha_i s_{iea}
 \ge v-G_e(v)/a-avt_{ea}.
$$
Since $av=\beta$, the resource constraint gives
$$
 z_e\ge\beta t_{ea}+\sum_i s_{iea}\ge v(1-G_e(v)/\beta).
$$

The maximum defining $Z_e$ is positive because $G_e(v)\to0$ as $v\downarrow0$.
Let $v_*>0$ attain it, and put $\tau_{\max,e}=\max_{i:e\in R_i}\tau_i$.
If $\tau_{\max,e}=0$, take $v_*=\Lambda_e$.
Otherwise, at every differentiability point,
$$
 \frac{d}{dv}[v(1-G_e(v)/\beta)]
 =1-G_e(v)/\beta-v\tau_e(v)/\beta\ge1-2v\tau_{\max,e}/\beta.
$$
Thus a maximizer satisfies $v_*\ge\min\{\Lambda_e,\beta/(2\tau_{\max,e})\}$.
It follows that $a_*=\beta/v_*$ lies between the chosen grid endpoints.
Round $a_*$ upward to the next grid scale $a'$.
Then $a_*\le a'<r a_*$, including when $a_*$ itself is a grid scale.
For $v'=\beta/a'$, we have $v_*/r<v'\le v_*$, and the grid-scale inequality gives
$$
 z_e\ge v'(1-G_e(v')/\beta)
 \ge v'(1-G_e(v_*)/\beta)
 >\frac{v_*}{r}(1-G_e(v_*)/\beta)=Z_e/r.
$$
Here $1-G_e(v_*)/\beta>0$ because $Z_e>0$.
\end{proof}

The next lemma extracts a useful affine lower bound from the same maximizer.

\begin{lemma}\label{lem:r2-empty-support}
Let $G=G_e$, $Z=Z_e$, and let $v_*>0$ maximize $v(1-G(v)/\beta)$.
There is $h=v_*/Z\in[1,2]$ such that, on $[0,\Lambda_e]$,
$$
 G(v)\ge\beta\max\left\{0,\frac{v-a}{Zh^2}\right\},
 \qquad a=Z(2h-h^2)\ge0.
$$
Consequently, for every $q>0$,
$$
 \int_0^{\Lambda_e}e^{-qG(v)}\,dv
 \le Z\left(2h-h^2+\frac{h^2}{q\beta}\right).
$$
\end{lemma}
\begin{proof}
Put $g=(\beta-G(v_*))/v_*=\beta Z/v_*^2>0$.
At an interior maximizer, the left and right derivatives of $v(1-G(v)/\beta)$ imply
$G'_-(v_*)\le g\le G'_+(v_*)$.
Thus $g$ is a supporting slope for the convex function $G$.
If $v_*=\Lambda_e$, the left derivative instead gives $g\ge G'_-(v_*)$, which is a valid supporting slope on the domain $[0,\Lambda_e]$.
Therefore
$$
 G(v)\ge G(v_*)+g(v-v_*).
$$
Setting $v=0$ and using $G(0)=0$ gives $G(v_*)\le g v_*=\beta-G(v_*)$.
Thus $0\le G(v_*)\le\beta/2$, and $h=1/(1-G(v_*)/\beta)\in[1,2]$.
The supporting line has slope $\beta/(Zh^2)$ and zero at $a=Z(2h-h^2)$.
Combining it with $G(v)\ge0$ proves the first assertion.
Extend this nonnegative affine bound beyond $\Lambda_e$ and integrate its exponential over $[0,\infty)$.
The result is $a+Zh^2/(q\beta)$, which proves the second assertion.
\end{proof}

\subsubsection{Exact purchase rates for nested clocks}

For a common $\theta\in[0,1)$, serve zero-$\tau_i$ types immediately and give every other type period
$$
 p_i(\theta)=s2^{\theta+\lceil\log_2\tau_i-\theta\rceil}.
$$
All clocks start at zero.
At a tick, clear exactly the due nonempty type queues and purchase their required resource union.
The periods are nested for every fixed $\theta$.
Let $p_\theta(\tau)$ denote this rounding function for $\tau>0$, with $p_\theta(0)=0$.
It is nondecreasing in $\tau$.
For resource $e$ define
$$
 P_{e,\theta}(v)=\int_0^v p_\theta(\tau_e(u))\,du,
 \qquad
 R_e(\theta)=\int_0^{\Lambda_e}e^{-P_{e,\theta}(v)}\,dv.
$$

\begin{lemma}\label{lem:r2-empty-exact-rate}
The expected number of purchases of resource $e$ through $T$ is at most $T R_e(\theta)$.
For every fixed phase and every $T\ge0$,
$$
 \mathbb E[\operatorname{ALG}_T]\le TQ(\theta)+K_I(\theta),
$$
where
$$
 Q(\theta)=\frac12\sum_i\lambda_i p_i(\theta)+\sum_e w_eR_e(\theta),
 \qquad
 K_I(\theta)=\sum_i C_i+\frac18\sum_i\lambda_i p_i(\theta)^2.
$$
\end{lemma}
\begin{proof}
Fix a resource and omit its subscript.
Let $\lambda_0$ be its total immediate rate and let $p_1<\cdots<p_m$ be its distinct positive periods, with corresponding group rates $\gamma_j$.
Put $B_0=0$ and $B_j=\sum_{k=1}^j\gamma_kp_k$.
A tick of period $p_j$ that is not a tick of $p_{j+1}$ has exactly groups $1,\ldots,j$ due.
For each such group, its pending count comprises its arrivals since its preceding own clock tick, an interval of length $p_k$.
The counts of distinct types are independent, so the probability of a nonempty clearing is $1-e^{-B_j}$.
These intervals lie in $[0,T]$ for every positive tick through $T$.
The nesting of the periods therefore gives the exact expectation
$$
 \lambda_0T+\sum_{j=1}^m
 \left(\left\lfloor\frac T{p_j}\right\rfloor-\left\lfloor\frac T{p_{j+1}}\right\rfloor\right)(1-e^{-B_j})
 =\lambda_0T+\sum_{j=1}^m\left\lfloor\frac T{p_j}\right\rfloor(e^{-B_{j-1}}-e^{-B_j}),
$$
where the final floor with $p_{m+1}$ is defined to be zero.
Immediate arrivals coincide with clock ticks only on a null event.
Every coefficient in the last sum is nonnegative, so the expectation is at most
$$
 T\left[\lambda_0+\sum_{j=1}^m\frac{e^{-B_{j-1}}-e^{-B_j}}{p_j}\right]
 =T\int_0^{\Lambda_e}e^{-P_{e,\theta}(v)}\,dv.
$$
The last identity follows by integrating over each constant-period rate group; the initial zero-period segment contributes $\lambda_0$.
It also covers $m=0$.

After $T$, each positive-period type has at most one final clearing, costing in total at most $\sum_i C_i$.
Its expected waiting cost is at most $\lambda_i(Tp_i/2+p_i^2/8)$.
Summing proves the claimed finite-horizon inequality, including completion.
\end{proof}

\subsubsection{Averaging the common logarithmic shift}

\begin{lemma}\label{lem:r2-empty-average}
For uniform $\theta\in[0,1)$,
$$
 \mathbb E_\theta[R_e(\theta)]\le\frac{Z_e}{1-1/(2s\beta\ell)}=\frac Kr Z_e,
 \qquad
 \mathbb E_\theta[Q(\theta)]<KL^*.
$$
\end{lemma}
\begin{proof}
For $\tau>0$, write $p_\theta(\tau)=s\tau U_\theta(\tau)$.
For every fixed positive $\tau$, the rounding multiplier $U_\theta(\tau)$ has the same distribution as $2^U$ with $U$ uniform on $[0,1)$.
The multipliers for different values of $\tau$ need not be independent.

Fix $v$ with $G_e(v)>0$.
Use the weights $\tau_e(u)/G_e(v)$ on $0<u<v$ and the convexity of the exponential to obtain
$$
 e^{-P_{e,\theta}(v)}
 \le\int_0^v\frac{\tau_e(u)}{G_e(v)}
 \exp\{-sG_e(v)U_\theta(\tau_e(u))\}\,du.
$$
Zero-$\tau_e(u)$ terms have zero weight and are omitted.
Averaging in $\theta$ uses only the identical one-dimensional multiplier distributions, yielding
$$
 \mathbb E_\theta[e^{-P_{e,\theta}(v)}]
 \le\,\mathbb E_U[e^{-s2^U G_e(v)}].
$$
If $G_e(v)=0$, then $P_{e,\theta}(v)=0$ and the same inequality is an equality.
Tonelli's theorem and Lemma~\ref{lem:r2-empty-support}, with the same $h$ for every $U$, now give
$$
 \mathbb E_\theta[R_e(\theta)]
 \le Z_e\left[2h-h^2+\frac{h^2}{s\beta}\,\mathbb E_U[2^{-U}]\right]
 =Z_e\left[2h-\left(1-\frac1{2s\beta\ell}\right)h^2\right].
$$
Here $\mathbb E_U[2^{-U}]=1/(2\ell)$.
Put $\delta=1/(2s\beta\ell)=D/(\beta K)<1$.
Completing the square gives $2h-(1-\delta)h^2\le1/(1-\delta)$ for every real $h$.
As $K=r+D/\beta$, this coefficient equals $K/r$.

The mean period is $\mathbb E_\theta[p_i]=s\tau_i/\ell$, so
$$
 \mathbb E_\theta[Q(\theta)]
 \le K\sum_i y_i+\frac Kr\sum_e w_eZ_e
 <K\left(\sum_i y_i+\sum_e w_ez_e\right)=KL^*.
$$
The strict inequality follows from Lemma~\ref{lem:r2-empty-prefix}.
\end{proof}

\paragraph{Selecting the phase.}
Test dyadic phases with successively finer certified evaluations and
accept the first one for which $Q(\theta)<KL^*$ is certified.
At round $m$, test the finite grid
$\{j2^{-m}:0\le j<2^m\}$ at precision $2^{-m}$, skipping any
candidate whose rounding decisions are unresolved at that precision.
This search uses only the instance and the dual scales $\tau_i$.
For evaluation, the preceding proof expresses every $R_e$ as a
finite sum of exponentials; no numerical integration is required.

\begin{lemma}\label{lem:r2-empty-deterministic}
The phase search terminates and returns $Q(\theta)<KL^*$.
\end{lemma}
\begin{proof}
Lemma~\ref{lem:r2-empty-average} gives
$\int_0^1Q(\theta)d\theta<KL^*$.
The only possible discontinuities are the finitely many rounding
breakpoints. Thus some open interval between them has
$Q(\theta)<KL^*$, with a positive margin on a smaller interval.
A dyadic phase lies in that smaller interval. At sufficiently high
precision its rounding decisions and the strict comparison are
certified, so the search stops. This proves finite termination,
without asserting an efficient bound on the preprocessing time.
\end{proof}

\begin{theorem}\label{thm:r2-empty-final}
For every fixed $r>1$ and $\beta>0$, the preceding preprocessing and nested-clock policy define a deterministic causal algorithm satisfying
$$
 \mathbb E[\operatorname{ALG}_T]\le
 \left(r+\frac1{4\beta\ln^22}\right)L^*T+K_I
 \qquad(T\ge0).
$$
If the same LP has a valid finite-horizon lower certificate
$$
 \mathbb E[\operatorname{OPT}_T]\ge\kappa L^*T-B_I
$$
with $\kappa>0$ and a fixed finite $B_I\ge0$, then
$$
 \mathbb E[\operatorname{ALG}_T]\le\frac{r+D/\beta}{\kappa}\,\mathbb E[\operatorname{OPT}_T]
 +K_I+\frac{r+D/\beta}{\kappa}B_I.
$$
In particular its long-run ratio is at most $(r+D/\beta)/\kappa$.
If the same positive coefficient $\kappa$ is valid for all the grids
obtained by varying $r>1$, then for every $\varepsilon>0$ a finite grid
gives a ratio at most $(1+D/\beta)/\kappa+\varepsilon$.
\end{theorem}
\begin{proof}
Use the phase from Lemma~\ref{lem:r2-empty-deterministic} in Lemma~\ref{lem:r2-empty-exact-rate}.
The second assertion follows by rearranging the stated lower certificate.
The completion constant is finite and depends only on the fixed instance.
On a nontrivial instance $L^*>0$, so it disappears in the long-run ratio.
For the final assertion take $r=1+\kappa\varepsilon$ and use the
assumed uniformity of $\kappa$ across these grids. The boundary constant
$B_I$ may depend on the chosen grid, since that grid is fixed before
the horizon tends to infinity.
No finite grid is claimed to attain the limiting coefficient itself.

\end{proof}

\begin{theorem}[Prescribed-resource guarantee]\label{ag:pair:thm:main}
For every $r>1$ and $\beta>0$, the finite preprocessing and nested
clocks above define a deterministic causal algorithm such that
$$
 \E[\ALG_T]\le \frac{r\beta+D}{C(\beta)}\,\E[\OPT_T]+K_I
 \qquad(T\ge0).
$$
In particular, $r=10001/10000$ and $\beta=27/20$ give ratio
$$
 C_R:=\frac{(10001/10000)(27/20)+D}{C(27/20)}<2.110250205.
$$
The same guarantee holds for rooted and prescribed-pair aggregation
on trees.
\end{theorem}
\begin{proof}
Combine Theorems~\ref{thm:r2-sharp-lp} and
\ref{thm:r2-empty-final}. Appendix~\ref{app:scalar-certificates}
certifies $C(27/20)>0.886377$. The positive series
$\ln2>2\sum_{j=0}^{39}[3^{2j+1}(2j+1)]^{-1}$ bounds $D$ from
above and gives the strict displayed numerical inequality by rational
arithmetic. On a nontrivial instance a positive feasible charge gives
$L^*>0$, so the offline lower bound is linear in $T$ and $K_I$
vanishes in the ratio. The path interpretations at the start of the
section prove the tree consequences.
\end{proof}

\subsection{Polynomial-time preprocessing}\label{sec:efficient-phase}
\paragraph{Input and guarantee.}
Assume rational costs and rates in binary, fix rational $r>1$ and $\beta>0$,
and let $0<\varepsilon\le1$ be rational. Write $B$ for the total
binary input length, including the encoding of $\varepsilon$.
The construction below takes time polynomial in that length and $1/\varepsilon$.
The all-free case is served immediately and bypasses the LP.
\paragraph{The LP grid.}

For a type $i$, put $C_i=\sum_{e\in R_i}w_e$, and let
\[
 \Gamma=\sum_i\lambda_i,\qquad
 \lambda_{\min}=\min_i\lambda_i,\qquad
 M=\sum_i\lambda_i C_i.
\]
The preceding grid contains consecutive powers of $r$, starting at
the last power at most $\beta/(2\Gamma)$ and ending at the first power
at least $\max\{2M/\lambda_{\min},\beta/\lambda_{\min}\}$.
For fixed rational $r,\beta$, the logarithmic magnitudes of these
endpoints are polynomial in $B$. Thus the number of grid points, the
absolute values of their exponents, and their rational encoding lengths
are polynomial in $B$. Exact comparisons of rational powers locate the
endpoints without computing logarithms.

The resource primal and dual therefore have polynomially many variables
and constraints with polynomial-size rational coefficients. Rational
linear programming gives optimal rational solutions of polynomial
encoding length in polynomial time. Let $L^*$ be their common value,
and set $\tau_i=y_i/\lambda_i$ for the selected optimal dual.

In a nontrivial instance, $L^*>0$. Indeed, choose an incidence $(i,e)$
and a scale $a$. Set $c_{ea}=w_e$ and
\[
 \delta_{iea}=\min\left\{w_e,a,\frac{\beta w_e}{a\lambda_i}\right\}>0,
\]
with all other variables zero. This is a feasible primal solution of
positive rational value. In particular, accuracy proportional to $L^*$
requires only polynomially many bits.

\subsubsection{A finite rational phase grid}

We use the preceding exact nested-clock rate bound. For a vector of
nested periods $p_i\ge0$, a zero period means immediate service. At
resource $e$, sort its users by period and let $p_e(v)$ be the resulting
rate-quantile function on $[0,\Lambda_e]$, where
$\Lambda_e=\sum_{i:e\in R_i}\lambda_i$. Put
\[
 P_e(v)=\int_0^v p_e(u)\,du,\quad
 R_e(p)=\int_0^{\Lambda_e}e^{-P_e(v)}\,dv,\quad
 Q(p)=\frac12\sum_i\lambda_i p_i+\sum_e w_eR_e(p).
\]
The manuscript proves
\begin{equation}\label{eq:clock}
 \E[\ALG_T]\le TQ(p)+\sum_i C_i+\frac18\sum_i\lambda_i p_i^2.
\end{equation}
These expressions retain the probability that a scheduled tick is empty.

\begin{lemma}[Common scaling]\label{lem:scaling}
For $0<c\le1$ and $\alpha\ge1$,
\[
 R_e(cp)\le c^{-1}R_e(p),\qquad R_e(\alpha p)\le R_e(p),
\]
and consequently
\[
 Q(cp)\le c^{-1}Q(p),\qquad Q(\alpha p)\le\alpha Q(p).
\]
\end{lemma}
\begin{proof}
The function $P_e$ is convex, nonnegative, and zero at the origin.
Thus $P_e(u)\le cP_e(u/c)$ for $0\le u\le c\Lambda_e$, including
when immediate types create an initial flat interval. Substitution gives
\[
 cR_e(cp)=\int_0^{c\Lambda_e}e^{-cP_e(u/c)}\,du
 \le\int_0^{c\Lambda_e}e^{-P_e(u)}\,du\le R_e(p).
\]
Increasing every period decreases each exponential integrand. Waiting
scales linearly, yielding the two inequalities for $Q$.
\end{proof}

Write $s=2K\ln2=2r\ln2+1/(2\beta\ln2)$ and replace the phase
$\theta\in[0,1)$ by $a=2^\theta\in[1,2)$. For $\tau_i>0$, define
\[
 k_i(a)=\min\{k\in\mathbb Z:a2^k\ge\tau_i\},\qquad
 p_i(a)=sa2^{k_i(a)}.
\]
For $\tau_i=0$, set $p_i(a)=0$. Each positive type has exactly one
possible breakpoint in this interval, namely the rational mantissa
\[
 b_i=\tau_i/2^{\lfloor\log_2\tau_i\rfloor}\in[1,2).
\]
The binary exponent and mantissa are computed by exact rational
comparisons. At a breakpoint the exponent drops; the value at the
breakpoint belongs to the cell to its right.

Set $\delta=\varepsilon/4$ and form
\[
 \mathcal H=\{1+j\delta:j\in\mathbb Z_{\ge0},\ 1+j\delta<2\}
                \cup\{b_i:\tau_i>0\}.
\]
There are $O(|I|+1/\varepsilon)$ candidates, all of polynomial encoding
length. For any $a_*\in[1,2)$, take its largest predecessor
$a\in\mathcal H$. If $a=a_*$ there is nothing to prove. Otherwise there
is no breakpoint in $(a,a_*]$, so all exponents agree and
\[
 p(a)=c\,p(a_*),\qquad
 c=\frac a{a_*}\ge\frac1{1+\delta}.
\]
The predecessor is in the same rounding cell even if it is itself a
breakpoint. No separation bound between distinct breakpoints is needed.
Lemma~\ref{lem:scaling} therefore gives
\begin{equation}\label{eq:phase-loss}
 Q(p(a))\le(1+\delta)Q(p(a_*)).
\end{equation}
Lemma~\ref{lem:r2-empty-average} supplies some $a_*$ with
$Q(p(a_*))\le KL^*$. We do not need a quantitative lower bound on the
strict margin used in the original phase search.

\subsubsection{Rational periods and certified candidate evaluation}

Choose rational $\bar s$ with $s\le\bar s\le(1+\delta)s$.
For example, rational bounds $l\le\ln2\le u$ give
\[
 2rl+\frac1{2\beta u}\le s\le
 2ru+\frac1{2\beta l}.
\]
The positive series
\[
 \ln2=2\sum_{j\ge0}\frac1{(2j+1)3^{2j+1}}
\]
has tail at most $(3/4)9^{-N}$ after $N$ terms. Refine these rational
bounds until their induced upper and lower bounds for $s$ have ratio
at most $1+\delta$, and use the upper bound as $\bar s$.
For fixed $r,\beta$, this takes $O(\log(1/\delta))$ terms.

At each candidate use
$\bar p_i(a)=\bar s\,a2^{k_i(a)}$, with zero for immediate types.
All periods are rational and nested. By common scaling and
\eqref{eq:phase-loss}, some candidate satisfies
\[
 Q(\bar p(a))\le(1+\delta)^2KL^*.
\]
Compute a certified interval $[\underline Q_a,\overline Q_a]$ of width
at most $\delta L^*$ for each candidate, and choose the smallest
upper endpoint. The chosen periods satisfy
\[
 Q(\bar p)\le (1+\delta)^2KL^*+\delta L^*
 \le\bigl[(1+\delta)^2+\delta\bigr]KL^*
 \le(1+\varepsilon)KL^*,
\]
because $K>1$ and $3\delta+\delta^2\le\varepsilon$ for
$0<\varepsilon\le1$.

To verify the evaluation complexity, group equal positive periods at
each resource as $p_1<\cdots<p_h$, with group rates $\gamma_j$, and put
$B_j=\sum_{k\le j}\gamma_kp_k$, $B_0=0$. Then
\[
 R_e=\lambda_{0,e}
       +\sum_{j=1}^h\frac{e^{-B_{j-1}}-e^{-B_j}}{p_j}.
\]
All arguments and coefficients are rational of polynomial encoding
length. Their total absolute coefficient size, together with the
positive rational $L^*$, determines a polynomial number of required
accuracy bits. Tiny exponential differences need not be evaluated to
small relative error.

For a nonnegative rational $x$, $e^{-x}$ can be enclosed to absolute
error $2^{-b}$ in polynomial time. If $x\ge b+1$, use
$0\le e^{-x}<2^{-b-1}$. Otherwise $x$ has polynomial magnitude;
Taylor's formula with its factorial remainder bound uses polynomially
many rational operations to reach the requested accuracy.
Interval operations then produce the required candidate bounds.

Finally, combine \eqref{eq:clock} with the preceding lower bound
$\E[\OPT_T]\ge[C(\beta)/\beta] L^*T$. Every positive-period type clears by its
next finite tick, proving completion and the following theorem. The displayed boundary constant remains valid
with the selected rational periods.

\begin{theorem}[Efficient prescribed-resource guarantee]\label{thm:efficient-resource}
For fixed rational $r>1$, $\beta>0$ and rational input, deterministic
preprocessing polynomial in the input bit length and $1/\varepsilon$
produces rational nested clocks satisfying
\[
 \E[\ALG_T]\le(1+\varepsilon)\frac{r\beta+D}{C(\beta)}
                  \E[\OPT_T]+K'_I.
\]
Here $K'_I$ includes all completion costs and is independent of $T$.
With the parameters of Theorem~\ref{ag:pair:thm:main} and
$\varepsilon=10^{-3}$, the ratio is below $2.113$.
\end{theorem}
\begin{proof}
The finite candidate construction gives
$Q\le(1+\varepsilon)KL^*$, where $K=r+D/\beta$.
Combine the exact clock bound with Theorem~\ref{thm:r2-sharp-lp}.
The selected rational periods are finite, so every finite input completes.
The preceding paragraphs establish polynomial encoding and computation.
\end{proof}

\paragraph{Rooted multi-level aggregation.}
Taking $R_v$ to be the path from a prescribed root to $v$ preserves
both online and offline costs on every realized input. Hence the same
$2.113$ efficient guarantee applies to stochastic rooted MLA, improving
on the general bound $210$ reported by Mari, Paw\l{}owski, Ren, and
Sankowski \cite{mari-et-al-2024}. This uses the new LP and clock policy
and does not depend on, or establish correctness of, their GEN analysis.
The running-time
claim is a polynomial bit-complexity statement, not a strongly polynomial
bound, and fixes the scale-grid parameters $r,\beta$.

\section{Concluding remarks}
Known Poisson rates permit a common schedule even when required resources
overlap arbitrarily. The resource LP allocates purchase and waiting
budgets without repeated charging; its dual coordinates the queues,
including the possibility that a clock tick is empty. A finite rational
phase grid makes preprocessing polynomial with an arbitrarily small
multiplicative loss. The same result gives the improved rooted MLA
bound by an exact path interpretation.

The competitive gap for general resources remains between
$1/C(1)>1.4082077033$ and $C_R<2.110250205$ with finite preprocessing.
The scalar family attains exactly $1/C(1)$, but this does not
determine the answer for arbitrary resource sets. Closing the gap
requires either a stronger algorithm for general overlap or an
instance whose optimal causal-to-clairvoyant ratio exceeds the
one-resource factor.

Further questions include exploiting tree structure beyond the generic
resource guarantee, replacing the large fixed grid constants by a more
practical optimization method, and learning unknown arrival rates.
The model fixes those rates throughout execution; nonstationary demand
requires a separate analysis.

\section*{Declaration of AI use}
Generative AI tools (OpenAI's Astra and Codex) assisted mathematical
exploration, proof development, drafting and revision, literature searches,
and LaTeX preparation. The authors are responsible for the correctness of
the results and references and for the final text.

\bibliographystyle{plainurl}
\bibliography{references}

\appendix
\section{Scalar arithmetic certificates}\label{app:scalar-certificates}
\subsection{Finite-state certificates for the exact scalar offline rate}

Let $C(q)$ denote the unrestricted offline cost per unit time for a
rate-one Poisson process with setup cost $q>0$ and additive waiting.
The cluster characterization in Section~\ref{sec:r2-single-exact}
also gives
$$
 C(q)=\lim_{n\to\infty}\frac{\mathbb E[D_n(q;X_1,\ldots,X_{n-1})]}{n},
 \qquad X_j\stackrel{\mathrm{iid}}\sim\operatorname{Exp}(1),
$$
where $D_n$ is the exact consecutive-batch dynamic program.
Indeed the regenerative clusters have geometric size $N$ and finite
mean cost $D$, and both the per-arrival and per-time rates equal
$\mathbb E[D]/\mathbb E[N]$ because the original arrival rate is one.
All comparisons below use this original per-arrival normalization.
We never divide by the mean of the quantized gaps.

\subsubsection{An exact finite state for integer gaps}

First take an integer setup cost $M\ge1$ and integer gaps in
$\{0,1,\ldots,M\}$.
After $n$ arrivals, a candidate final batch containing $j$ requests has
an excess cost $d$ above the current optimum $D_n$.
Retain only pairs $(j,d)$ with $0\le d<M$.
A pair is discarded if another has no larger $j$ and no larger $d$.
When equal first coordinates occur, keep only the smaller excess.
Store the remaining pairs with increasing $j$ and strictly decreasing
$d$.
At least one retained pair has excess zero.

An excess $d\ge M$ is never needed again.
At the next arrival, starting a singleton costs $D_n+M$, while this
old candidate costs at least $D_n+d$ even if the gap is zero.
The singleton also has a smaller batch size, so the old candidate
remains dominated after any further extension.
The same comparison proves the Pareto deletion rule.

Every pending count $j\ge M$ can be replaced by $M$.
While subsequent gaps are zero, its magnitude has no effect on cost.
At the first positive integer gap, such a candidate increases its
cost by at least $M$ and is dominated by a new singleton.
If equality occurs, the singleton has the smaller pending count.
Thus this truncation remains valid even when arbitrarily many
consecutive quantized gaps are zero.
Equal truncated counts are again merged by taking the least excess.

For a new gap $k$, the exact marginal cost is
$$
 G(s,k)=\min\left\{M,\min_{(j,d)\in s}(d+jk)\right\}.
$$
Replace each old pair by
$$
 \left(\min\{M,j+1\},\ d+jk-G(s,k)\right),
$$
add the new singleton $(1,M-G(s,k))$, and apply the deletion and
merging rules above.
Pairs with excess $M$ are deleted.
This gives a deterministic transition $s\mapsto T(s,k)$.
The initial state after the first request is $s_0=\{(1,0)\}$.
The stored coordinates lie in
$\{1,\ldots,M\}\times\{0,\ldots,M-1\}$, so there are finitely many
possible states.
Breadth-first search enumerates precisely the states reachable from
$s_0$ under the chosen alphabet.
No queue-size restriction is imposed on the original instance.

If the gap $M$ has positive probability, it sends every state to
$s_0$.
The resulting finite Markov chain therefore has a unique limiting
distribution: two copies driven by the same gaps coalesce at their
first occurrence of $M$.
If $P$ is its stochastic transition matrix and $r(s)$ its expected
one-step marginal cost, the exact long-run cost per arrival is
$$
 \gamma=\pi r,
$$
where $\pi P=\pi$.
This also follows directly by telescoping
$D_n=M+\sum_{j=1}^{n-1}G(s_j,k_j)$ and taking expected Cesaro averages.

\subsubsection{Lower and upper quantizations of exponential gaps}

In the continuous scalar problem, every gap at least $q$ can be made
an optimal batch boundary.
Consequently replacing $X$ by $\min\{X,q\}$ leaves every finite
offline dynamic-programming value unchanged.
Let $b$ be the number of bins, put $h=q/b$, and define
$$
 a_j=e^{-jh}-e^{-(j+1)h},\qquad
 \beta_h=1-\frac{he^{-h}}{1-e^{-h}},\qquad \theta_h=\beta_h/h.
$$
The conditional mean within bin $[jh,(j+1)h]$ is $jh+\beta_h$.

For the lower quantization, given a capped gap $X=x$ in
$[jh,(j+1)h]$, set $Y=(j+1)h$ with probability $(x-jh)/h$
and set $Y=jh$ with the remaining probability.
Use independent rounding coins conditional on the original gaps.
Thus $\mathbb E[Y\mid X]=X$ for each original capped gap, not merely
after averaging over its bin.
For the atom $X=q$, set $Y=q$.
The resulting probabilities on the grid $\{0,h,\ldots,bh\}$ are
$$
 \begin{aligned}
 p_0^-&=a_0(1-\theta_h),\\
 p_j^-&=a_{j-1}\theta_h+a_j(1-\theta_h)\quad(1\le j<b),\\
 p_b^-&=a_{b-1}\theta_h+e^{-q}.
 \end{aligned}
$$
The final term includes the entire atom of the capped gap at $q$.
The setup cost becomes $M=b$ in units of $h$, and the gap alphabet
is $\{0,1,\ldots,b\}$.

For the upper quantization, condition on the bins and replace each
finite-bin gap by its midpoint.
The exponential density is decreasing, so
$\beta_h\le h/2$.
This follows, for example, by pairing $x$ and $h-x$ in the integral
of $(x-h/2)e^{-x}$ over $[0,h]$.
The integer setup cost is now $M=2b$ in units of $h/2$.
The gap alphabet is $\{1,3,\ldots,2b-1,2b\}$ with probabilities
$a_0,a_1,\ldots,a_{b-1},e^{-q}$.

The dynamic-programming value $D_n$ is the minimum of finitely many
affine functions of the gaps.
It is jointly concave and is nondecreasing in each gap.
For every fixed $n$, conditional Jensen therefore gives
$$
 \mathbb E[D_n(\text{lower grid})]
 \le\,\mathbb E[D_n(\text{original gaps})]
 \le\,\mathbb E[D_n(\text{upper grid})].
$$
The first inequality conditions on the original capped gaps and uses
the mean-preserving endpoint randomization.
The second conditions on the bins, applies Jensen at the conditional
means, and then increases each mean to its midpoint.
Dividing by $n$ and taking limits gives
$$
 \gamma_-(q,b)\le C(q)\le\gamma_+(q,b).
$$
These are comparisons of the unrestricted offline dynamic programs.
The artificial grids are used only to bound costs; they do not define
new physical arrival rates or restrict the original online policies.

\subsubsection{A finite arithmetic certificate for the stationary reward}

One need not approximate a stationary distribution.
For any stochastic matrix $P$ on the enumerated states and its reward
vector $r$, put $v_N=P^Nr$.
For every integer $N\ge0$,
$$
 \min_s v_N(s)\le\pi r\le\max_s v_N(s),
$$
because $\pi v_N=\pi r$.
This inequality is exact at every fixed iteration count.
Mixing rates are unnecessary for its validity.

The finite arithmetic certificate encloses every gap probability by exact rational
numbers, then rounds them outward to multiples of $10^{-14}$.
The lower probabilities need not sum to one, and neither do the upper
probabilities.
This causes no problem: all rewards and iterates are nonnegative.
If $\underline P\le P$ and $\underline r\le r$ entrywise, then
$\underline P^N\underline r\le P^Nr$.
An analogous inequality holds for upper bounds.
Every multiplication and addition is rounded downward for lower
bounds and upward for upper bounds.
Consequently the minimum of its lower iterate is a rigorous lower
bound on $\gamma$, and the maximum of its upper iterate is a rigorous
upper bound.

All probability enclosures are produced using exact fractions.
For $0\le x\le3/2$, the alternating
Taylor tails give
$$
 \sum_{j=0}^{41}\frac{(-x)^j}{j!}
 \le e^{-x}\le
 \sum_{j=0}^{40}\frac{(-x)^j}{j!}.
$$
The terms are decreasing in magnitude throughout the relevant tail.
The expression $1-hz/(1-z)$ is decreasing in $z$, which fixes the
direction of the conditional-mean bounds.
Products involving $\theta_h$ and $1-\theta_h$ are enclosed separately,
so no independence or common extremizer of these interval endpoints
is assumed.

\subsubsection{Certified values}
Both quantizations use $b=16$ bins. The endpoint graph has $59{,}666$
states and $1{,}014{,}322$ state--symbol transitions; the midpoint
graph has $18{,}666$ states and $317{,}322$ transitions. Each has
17 input symbols. Using 100 directed iterations gives the following
strict lower bounds. The graph is the same at each setup price;
only the probability enclosures and cost scale change.
$$
\begin{array}{c|r}
q&\text{lower bound on }C(q)\\\hline
7/8&0.6407995521894\\
1&0.709707\\
27/20&0.886377\\
189/100&1.1245989704271\\
3&1.533777646201\\
6&2.375717698470\\
10&3.218954343659
\end{array}
$$
The midpoint upper graph gives
$M_{16}<0.7102999314262$ at setup one. This is a bound on the
midpoint model; the next section corrects it to bound the original
continuous-gap optimum more sharply. For arguments up to ten the
probability generator uses alternating orders 80 and 81; the terms
decrease in magnitude throughout the relevant tails. The source
archive contains the graphs, exact probability generators, directed
evaluator, and raw output for every displayed value.

The bounds at $3,6,10$ imply the rational comparisons used in
Theorem~\ref{thm:r4-scalar-worst}. For the resource upper bound,
substitution of $C(27/20)>0.886377$ and the positive rational lower
series for $\ln2$ gives $C_R<2.110250205$.

\subsection{Conditional gap means and short clusters}
\label{sec:r6-scalar-correction}

We now use the conditional means inside the midpoint bins to improve the offline upper certificate.

Let $C(q)$ be the long-run offline cost per unit time of rate-one
Poisson requests, unit additive delay, and a setup cost $q>0$.
For $n$ consecutive requests, write $D_n(q;x)$ for their optimal
completion cost when their $n-1$ interarrival gaps are $x$.
As proved in the scalar analysis, an optimum consists of consecutive
batches served at their last arrival. Thus $D_n$ is the minimum of
finitely many affine functions of the gap vector, and is concave
and coordinatewise nondecreasing in that vector. Also
$\lim_{n\to\infty}\,\mathbb E[D_n(q;X)]/n=C(q)$ for independent
rate-one exponential gaps $X$.

\subsubsection{A finite-input correction of midpoint rounding}

Fix $m\ge1$, put $h=1/m$, and cap gaps at one.
On each interval $[jh,(j+1)h)$, $0\le j<m$, round a gap to its
midpoint $(j+1/2)h$; map every gap at least one to one.
Denote the resulting random gap vector by $Y$.
Throughout this section the distribution of $Y$ is fixed using this
setup-one grid, even when $D_n(q;Y)$ is evaluated at a setup cost
$q<1$.

Conditional on an internal bin, the mean of an exponential gap is
$$
 jh+1-\frac{h}{e^h-1}
 =(j+1/2)h-\delta,
 \qquad
 \delta=\frac h2-1+\frac{h}{e^h-1}>0.
$$
Let $\overline Y$ be $Y-\delta$ in each internal bin and one on
the capped tail bin. For every $0<q\le1$, capping a gap at one
does not change the optimum: an optimal schedule may cut at every
gap at least $q$. Conditional Jensen and monotonicity therefore give
$$
 \mathbb E[D_n(q;X)]
 \le\,\mathbb E[D_n(q;\overline Y)]
 \le\,\mathbb E[D_n(q;Y)].
$$

\begin{lemma}[Mean correction using a second setup price]
Fix $0<q<1$, and put
$$
 A_n=\mathbb E[D_n(1;Y)],\qquad
 F_n(q)=\mathbb E[D_n(q;X)].
$$
Then
$$
 F_n(1)\le A_n-\delta
       \left[n-\frac{A_n-F_n(q)}{1-q}\right].
$$
\end{lemma}
\begin{proof}
For each midpoint gap vector $Y$, choose an optimal setup-one
partition with $K_n$ batches that cuts at every gap equal to one.
Such a partition exists: splitting a batch at a gap of length one
adds one setup but saves at least one unit of delay.
Every uncut gap consequently belongs to an internal bin.
Replacing $Y$ by $\overline Y$ reduces the delay of this partition
by $\delta$ times the number of requests preceding each uncut gap
within its batch. There are $n-K_n$ uncut gaps, each with coefficient
at least one. Hence
$$
 D_n(1;\overline Y)
 \le D_n(1;Y)-\delta(n-K_n).
$$
The same partition, evaluated at setup price $q$, also gives
$$
 D_n(q;Y)\le D_n(1;Y)-(1-q)K_n,
 \qquad
 K_n\le\frac{D_n(1;Y)-D_n(q;Y)}{1-q}.
$$
These two inequalities use the same fixed midpoint gap vector.
Take expectations, use conditional Jensen for the setup-one cost,
and use $\mathbb E[D_n(q;Y)]\ge F_n(q)$.
The displayed claim follows.
\end{proof}

Let $M_m=\lim_{n\to\infty}A_n/n$ be the cost rate of the midpoint
model. The finite-state midpoint certificate already establishes
this limit. Taking limits in the preceding finite-input inequality
gives
$$
 C(1)\le M_m-\delta
       \left[1-\frac{M_m-C(q)}{1-q}\right].
$$
No derivative of $C$, interchange of perturbation limits, or
numerical mixing-time estimate is used here.
If $M_m\le U$ and $C(q)\ge L$, this implies
$$
 C(1)\le U-\delta\eta,
 \qquad
 \eta=1-\frac{U-L}{1-q}.
$$
When $\eta>0$, a certified lower bound on $\delta$ suffices for
a strict upper certificate on $C(1)$.

\subsubsection{An additional nonnegative charge from short clusters}

The exact coefficient of the gap shift for a partition with batch
sizes $\ell_1,\ldots,\ell_K$ is
$$
 W=\sum_j\frac{\ell_j(\ell_j-1)}2
   =(n-K)+\sum_j\frac{(\ell_j-1)(\ell_j-2)}2.
$$
The last sum is nonnegative. We can account for some of it without
computing longer clusters or changing the finite-state graph.
In the midpoint model, cut at every gap equal to one and call the
resulting groups clusters. Within each cluster choose a minimum-cost
partition, breaking ties by maximizing
$$
 E=\sum_j(\ell_j-1)(\ell_j-2)/2.
$$
These choices concatenate to a global minimum-cost partition.
The earlier setup-price comparison holds for every minimum-cost
partition, so this tie rule is compatible with the bound on $K$.

Write $p=e^{-1}$ and
$p_j=e^{-jh}-e^{-(j+1)h}$ for $0\le j<m$.
For a cluster of size $s$, let $E_s(j_1,\ldots,j_{s-1})$ be the
largest value of $E$ among optimal partitions with midpoint gaps
$(j_i+1/2)h$. For every fixed cutoff $s_*\ge3$, define
$$
 \Gamma_{s_*}
 =p^2\sum_{s=3}^{s_*}\ \sum_{j_1,\ldots,j_{s-1}=0}^{m-1}
      E_s(j_1,\ldots,j_{s-1})\prod_{i=1}^{s-1}p_{j_i}.
$$
The two factors $p$ impose the preceding and following cluster
boundaries. Thus $\Gamma_{s_*}$ is the expected retained extra charge
per arrival in a stationary sequence of gaps. More explicitly, in
an input of $n$ requests, each fixed size $s\le s_*$ has
$(n-s-1)_+$ possible cluster starts whose two boundary gaps are
inside the input. Linearity of expectation gives a total expected
extra charge at least $(n-s_*-1)_+\Gamma_{s_*}$.
Clusters larger than $s_*$ have nonnegative extra charge and are
simply omitted. No bound on their tail probabilities is needed.

Using this sharper coefficient in the same finite-input comparison
and then taking $n\to\infty$ proves
$$
 C(1)\le M_m-\delta
       \left[1-\frac{M_m-C(q)}{1-q}+\Gamma_{s_*}\right].
$$
Equivalently, this is a feasible clairvoyant witness: choose the
midpoint-optimal partition from the gap bins and serve each batch at
its last actual arrival. Conditional on the bins, its expected cost
is the cost with the conditional mean gaps, with the exact reduction
$\delta W$. It completes every finite input.

\subsubsection{A reproducible rational certificate}

Use $m=16$ and $q=7/8$. The midpoint graph
has $18{,}666$ states and gives the outward-rounded upper bound
$$
 U=0.7102999314262.
$$
The endpoint-rounding lower graph has $59{,}666$ states.
Evaluating that same graph with the exponential-bin probabilities
for setup $q=7/8$ gives the conservative lower bound
$$
 L=0.6407995521894.
$$
This lower certificate uses its own grid with spacing $q/16$.
That grid is used only to bound the continuous quantity $C(q)$;
it is not identified with the fixed setup-one midpoint distribution
in the finite-input lemma.

For a completely rational bound on $\delta$, put $h=1/16$ and
$$
 E_+=\sum_{j=0}^{20}\frac{h^j}{j!}
       +\frac{h^{21}}{21!}\frac1{1-h/22}.
$$
The positive exponential series gives $e^h<E_+$, so
$$
 \delta>\delta_-:=\frac h2-1+\frac{h}{E_+-1}.
$$
All quantities in the following comparison are rational:
$$
 \eta=1-8(U-L)=0.4439969661056>0,
 \qquad
 C(1)<U-\delta_-\eta<0.710155410573,
$$
before accounting for the additional short-cluster charge.
For the additional charge, exact integer dynamic programming over
all midpoint gap words with cluster size at most six gives
$$
 \Gamma_6>\Gamma_0:=0.101052891414.
$$
The four word counts for cluster sizes $3,4,5,6$ are respectively
$256$, $4096$, $65536$, and $1048576$.
Rescaling to setup cost $32$ gives the integer midpoint
gaps $1,3,\ldots,31$, so all dynamic-programming comparisons and
tie decisions are exact. Each exponential-bin probability and $p$
is enclosed using exact rational arithmetic. The nonnegative weighted
sum is evaluated with downward rounding after every multiplication
and addition, and its binary endpoint is converted to an exact fraction.
Consequently the stronger rational certificate is
$$
 C(1)<U-\delta_-(\eta+\Gamma_0)<0.710122517893,
$$
and
$$
 \frac1{U-\delta_-(\eta+\Gamma_0)}>1.4082077033.
$$
These strict comparisons are finite exact-rational calculations using
the probability enclosures and directed iterations specified above.
The ancillary files accompanying this arXiv version contain the
verification programs, finite-state graphs, probability bounds, and
complete arithmetic output. The file \texttt{certificates/README.txt}
within the ancillary material gives the dependencies and commands
needed to reproduce the numerical certificates. The programs verify
the finite arithmetic bounds in this appendix; the probabilistic and
algorithmic arguments are proved in the paper.

\end{document}